\documentclass[11pt,a4paper]{article}
\pdfoutput=1
\usepackage{fullpage}
\usepackage{graphicx}
\usepackage[applemac]{inputenc}
\usepackage{amssymb}

\begin{document}

\date{\today}

\title{Higgs boson decay into two photons in an electromagnetic background field\footnote{
Preprint Numbers: CP3-Origins-2013-50 DNRF90 and DIAS-2013-50
}}

\author{N. K. Nielsen,
\footnote{email: nkn@cp3-origins.net}\\
Center of Cosmology and Particle Physics Phenomenology (CP3-Origins),\\
University of Southern Denmark, \\
DK 5230 Odense M, Denmark}

\maketitle

\begin{abstract}
The amplitude for Higgs boson decay into two photons in a homogeneous and time-independent magnetic field is investigated by proper-time regularization in a gauge invariant manner and is found to be singular at large field values. The  singularity is caused by the component of the charged vector boson field that is tachyonic in a strong magnetic field. Also tools for the computation of the amplitude in a more general electromagnetic background are developed.\\
 {\em PACS  numbers 11.15.-q, 12.15.-y, 12.15.Lk, 14.70. Fm.}\\

\end{abstract}

\newpage

\section{Introduction}

Soon after the discovery of the 126 {\it GeV} Higgs boson  \cite{ATLAS}, \cite{CMS} it was pointed out by Olesen \cite{Olesen} (cf. also \cite{Tuchin})  that a large magnetic field is  generated by the quarks producing the Higgs boson,  and that this magnetic field might influence the decay processes of the Higgs boson, and in particular the decay $H \rightarrow\gamma \gamma$ (a Higgs boson decaying to two photons).

In the present paper it is proven that this  indeed is the case. The amplitude for this decay process is considered for the unrealistic case of a stationary homogeneous magnetic field $B$ by the method of Schwinger \cite{Schwinger}, further developed by Adler \cite{Adler} and by Tsai and Erber \cite{Tsai}. It is demonstrated that
the amplitude  contains a term proportional to 
\begin{equation}
\frac{ eB }{M_H^3\sqrt{M_W^2- eB -\frac{1}{4}M_H^2}}
\label{manderley}
\end{equation}
 (with $eB>0$) for emission of photons along the field lines, with $e$ the fundamental electric charge unit, $M_W$ the {\it W}-boson mass and $M_H$ the Higgs boson mass. The amplitude is thus singular at
$B=\frac 1e(M_W^2-\frac{1}{4}M_H^2)< B_{\rm crit}$, where $B_{\rm crit}=\frac{M_W^2}{e}$ is the critical field strength where a component of the $W$-field becomes tachyonic \cite{NO}, \cite{JO}. The singularity is caused by this would-be tachyonic field component (in agreement with Olesen's prediction \cite{Olesen}) and also by the fact that charged particles only propagate along the field lines, such that their loop Feynman integrals are effectively two-dimensional. The amplitude is exponentially damped for emission of photons   not aligned with the magnetic field, and the  denominator is modified in this case.

The amplitude of Higgs boson decay to two photons was first computed many years ago by Ellis, Gaillard and Nanopoulos \cite{Ellis} (see also \cite{Ioffe}-\cite{Marciano}). The influence of a background field on the amplitude has not been considered before, but the pioneering paper by Vanyashin and Terentev \cite{Vanyashin} dealing with the Heisenberg-Euler effective action caused by a charged vector field makes it possible to find the behavior of the amplitude in the limit where the photon energies  are close to zero, which is only possible with a    Higgs boson mass also close to zero. The result described above deals with a more general situation, and the factor $\frac{1}{M_H^3}$ makes a direct comparison difficult.  It turns out that the singularity of (\ref{manderley}) can not be found from the Heisenberg-Euler effective action. 

An issue relevant for the calculation is that of gauge parameter independence, where it recently was shown that  the $H \rightarrow\gamma \gamma$   amplitude is the same in all $R_\xi$-gauges \cite{Marciano}. This statement can be extended to a general electromagnetic background field, using  methods developed in a recent publication \cite{NKNny}, but the proof is omitted here because of its excessive length \footnote{It was included in an earlier version of this paper.}. It is plausible that a background field does not upset the proof of gauge parameter independence since the leading singularities of propagators at short distances are independent of the background field. In general one expects gauge parameter independence of the amplitude in a regularization scheme respecting BRST invariance  (this can be seen from \cite{NKNold}, sec. 4,  and also from \cite{Sirlin}). With this justification a particular gauge (the Feynman gauge) is used throughout this paper.

The layout of the paper is as follows: In sec. 2 the standard electroweak theory is recapitulated and  used to formulate an effective action at one-loop order describing Higgs boson decay to two photons in a background electromagnetic field. Formal developments in this construction are dealt with at length in app. A. It is also demonstrated in sec. 2 how the decay amplitude obtained by dimensional regularization   is found from the effective action by the proper-time method, and a heuristic argument is given for (\ref{manderley}).

Sections 3 and 4 constitute the central part of the paper. Sec. 3 contains a derivation of the decay amplitude in a general homogeneous field by the methods of \cite{Schwinger}, \cite{Adler}, \cite{Tsai}, while the singular terms in a homogeneous magnetic field are extracted from the amplitude in sec. 4. App. B contains material  on propagators and the associated kernels relevant for the following sections in the context of proper-time regularization. In app. C  it is proven that  the amplitude as well as its singular terms are invariant under gauge transformations of the radiation field. App. D. gives details on the connection to the Heisenberg-Euler effective action \cite{Vanyashin}.

 Finally quark contributions to the amplitude are considered in sec. 5 and found not to give rise to singularities induced by the magnetic field, while the Higgs boson self energy is shown in sec. 6 to possess a  singularity similar to (\ref{manderley}). 

\section{Electroweak theory and $H\rightarrow \gamma \gamma$ decay effective action}

\subsection{Electroweak theory}

The metric is $\eta_{\mu\nu}=(+---).$ 

In the standard electroweak theory the scalar Lagrangian is, keeping only terms relevant for Higgs boson decay to photons, with the Higgs boson field denoted $H$,  the charged Goldstone boson fields $\chi^\pm$ and charged vector boson fields $W^\pm_\mu$:
\begin{eqnarray}&&
{\cal L}_{\rm sc}=\frac 12 (\partial _\mu H
+\frac g2(W^-_\mu\chi ^++W^+_\mu \chi^-) )^2
+  (\chi^+\stackrel{\leftarrow}{D}^\mu -\frac g2W^{+\mu} H  ) (D _\mu \chi^- -\frac g2W^-_\mu H  ) 
\nonumber\\&&-\frac 12 \mu ^2 (2\chi ^+\chi ^-+H^2)
-\frac{\lambda }{4}(2\chi ^+\chi ^-+H^2)^2
\label{kagemucha}
\end{eqnarray}
with the coupling constants  $g$.
By the Higgs mechanism one makes the replacement $H\rightarrow v+H, v=\sqrt{\frac{-\mu^2}{\lambda}}$, and $W^\pm$ get the mass $M_W=\frac{gv}{2}$, while the Higgs boson mass is $M_H=\sqrt{2\lambda}v$.  The covariant derivatives are:
\begin{equation}
D_\mu =\partial _\mu -ie A_\mu, \stackrel{\leftarrow}{D}_\mu=\stackrel{\leftarrow}{\partial}_\mu+ieA_\mu
\end{equation}
with $e=g\sin \theta_W$ the elementary charge unit, where $\theta_W$ is  the Weinberg angle,  and with $A_\mu$ the electromagnetic field.

In order to describe radiation processes one  splits the electromagnetic field $A_\mu$:
\begin{equation}
A_\mu \rightarrow A_\mu +{\cal A}_\mu
\label{sossamon}
\end{equation}
with $A_\mu$ a background field,  and ${\cal A}_\mu$ the radiation field, which fulfils the wave equation and has two independent transverse polarizations.  The interaction between radiation and $W$-bosons is described by  the action:
$$
-\int d^4x W^{+\nu}{\cal H}_{\nu\mu}W^{-\mu}$$
with ${\cal H}$  given by:
\begin{eqnarray}&&
{\cal H}_{\nu \mu}=-2ie{\cal F}_{\mu \nu}+2ie\eta _{\mu \nu}{\cal A}^\lambda D_\lambda +ie(\stackrel{\leftarrow}{D}_\nu{\cal A}_\mu -{\cal A}_\nu D_\mu)-e^2({\cal A}_\mu {\cal A}_\nu-\eta _{\mu \nu}{\cal A}^\lambda {\cal A}_\lambda)
\nonumber\\&&
={\cal H}^{(1)}_{\nu \mu}+{\cal H}^{[2]}_{\nu \mu}
\label{supergill}
\end{eqnarray}
where  the superscript denotes the order in $e$ and where we introduced the radiation field strength:  
\begin{equation}
{\cal F}_{\mu \nu}=\partial_\mu {\cal A}_\nu-\partial_\nu {\cal A}_\mu. 
\label{tajmahal}
\end{equation}
The following relations followis from (\ref{supergill}) and the on-shell properties of the radiation  field ${\cal A}_\mu $:
\begin{eqnarray}&&
D^\nu  {\cal H}^{(1)}_{\nu \mu}=-ie{\cal A}^\nu(\eta _{\nu \mu}D^2+ D_\nu D_\mu-2D_\mu D_\nu),
\nonumber\\&&
  {\cal H}^{(1)}_{\nu \mu}\stackrel{\leftarrow}{D}^\mu=ie(\eta _{\nu \mu}\stackrel{\leftarrow}{D}^2+ \stackrel{\leftarrow}{D}_\nu \stackrel{\leftarrow}{D}_\mu-2\stackrel{\leftarrow}{D}_\mu \stackrel{\leftarrow}{D}_\nu){\cal A}^\mu
\label{ehuduanna}
\end{eqnarray}
which should be understood as  relations between differential  operators.

The gauge of $W^\pm$ is fixed by:
\begin{eqnarray}&&
\mathcal{L}_{\rm gf}  =- ( W^{+,\mu}\stackrel{\leftarrow}{D}_\mu+\frac{gv}{2}\chi^+)(D_\nu W^{-,\nu}+\frac{gv}{2}\chi^-)
\label{hakonjarl}
\end{eqnarray}
(the $R_\xi$ Feynman gauge).
By  (\ref{hakonjarl})  a Goldstone boson mass squared $ M_W^2$ is generated. The Faddeev-Popov ghost Lagrangian is:
\begin{equation}
{\cal L}_{\rm FP} = - \bar{c}^+(c^+\stackrel{\leftarrow}{D}^2+ie({\cal A}^{\mu}c^+)\stackrel{\leftarrow}{D}_\mu+\frac{g^2v}{4}Hc^+)
- \bar{c}^-(D^2c^--ieD_\mu({\cal A}^{\mu}c^-)+\frac{g^2v}{4}Hc^-)
\label{ghostafson}
\end{equation}
so the ghost mass is equal to the  Goldstone boson mass.

\subsection{Proper-time representation of the scalar and vector propagators in a general background}

The scalar propagator $G_{\rm sc }(x,x')$ corresponding to the mass $ M_W^2$ is given by:
\begin{equation}
G_{{\rm sc}}(x,x')=\int _0^\infty d\tau h_{{\rm sc} }(x, x';\tau)
\label{ivantaurus}
\end{equation}
 with $D^2=\eta ^{\mu \nu}D_\mu D_\nu$ and with $\tau$ the proper time variable \cite{Schwinger}, \cite{DeWitt}, and:
\begin{eqnarray}&&
(D^2+ M_W^2)G_{{\rm sc}}(x,x')=G_{{\rm sc}}(x,x')(\stackrel{\leftarrow}{D'}^2+ M_W^2)=-i\delta(x-x')
\label{carmoisin}
\end{eqnarray}
where a primed derivative refers to $x'$ and where the scalar kernel $h_{{\rm sc}}(x, x';\tau)$ is defined by:
\begin{equation}
(i\frac{\partial}{\partial \tau}-(D^2+ M_W^2)) h_{{\rm sc}}(x, x';\tau)=0, \   h_{{\rm sc} }(x, y;0)=\delta(x-x').
\label{kenneweg}
\end{equation} 

 The   vector propagator  $G_{{\rm vec}, \mu \nu}(x, x')$ is similarly defined by:
\begin{eqnarray}&&
(D^2+M_W^2)G_{{\rm vec},  \mu \nu}(x, x')-2ieF_{\mu \lambda}(x)G_{{\rm vec}, }\hspace{-0.01 mm}^\lambda \hspace{0.1 mm}_ \nu(x, x')
\nonumber\\&&
=
G_{{\rm vec},  \mu \nu}(x, x')(\stackrel{\leftarrow}{D'}^2+M_W^2)-G_{{\rm vec}, \mu}\hspace{-0.01 mm}^\lambda (x, x')2ieF_{\lambda \nu }(x')
\nonumber\\&&
=i\eta_{\mu \nu}\delta(x, x')
\label{magenta}
\end{eqnarray}
where  $F_{\mu \nu}=\partial_\mu A_\nu-\partial_\nu A_\mu$ is the background field strength. 
The solution of (\ref{magenta}) is:
\begin{equation}
G_{{\rm vec},  \mu \nu}(x, x')=\int _0^\infty d\tau h_{{\rm vec},  \mu \nu}(x, x';\tau)
\label{timian}
\end{equation}
with:
\begin{eqnarray}&&
(i\frac{\partial}{\partial \tau}-(D^2+M_W^2))h_{{\rm vec},  \mu \nu}(x, x';\tau)+2ieF_{\mu}\hspace{0.1 mm}^\lambda h_{{\rm vec}, \lambda \nu}(x, x';\tau)=0, 
\nonumber\\&&
h_{{\rm vec},  \mu \nu}(x, x';0)=-\eta_{\mu\nu}\delta(x-x')
 \label{ginsberg}
\end{eqnarray}
defining the vector kernel corresponding to the scalar kernel defined by (\ref{kenneweg}). The integration path in (\ref{ivantaurus}) and (\ref{timian}) can be deformed such that it runs below the real axis or along the negative imaginary axis in the complex $\tau$-plane, provided no field components are tachyonic.

The following Ward identities hold for the kernels:
\begin{eqnarray}&&
D^\mu h_{{\rm vec},   \mu \nu}(x, x';\tau )=h_{{\rm sc}}(x, x';\tau)\stackrel{\leftarrow}{D}_\nu,
\nonumber\\&&
h_{{\rm vec},  \mu \nu}(x, x';\tau)\stackrel{\leftarrow}{D}^\nu=D_\mu h_{{\rm sc}}(x, x';\tau)
\label{draW}
\end{eqnarray}
since both sides of the two equations obey the same first-order differential equations in $\tau$ with the same boundary conditions; here was also used:
\begin{equation}
D^\nu D^2-D^2D^\nu=-2ie F^{\nu \lambda}D_\lambda .
\label{ronkedor}
\end{equation}
following from  the defintion of the covariant derivative and the fact that the background field is a solution of the Maxwell equations. From  (\ref{draW}) follows the Ward identities of propagators:
\begin{eqnarray}&&
D^\mu G_{{\rm vec},  \mu \nu}(x, x')=G_{{\rm sc}}(x, x')\stackrel{\leftarrow}{D}_\nu,
\nonumber\\&&
G_{{\rm vec},  \mu \nu}(x, x')\stackrel{\leftarrow}{D}_\nu=D_\mu G_{{\rm sc}}(x, x').
\label{Ward}
\end{eqnarray}

\subsection{$H\rightarrow \gamma \gamma$ decay effective action}

A background Higgs boson  field $H(x)$  is used here which is on-shell, i.e. 
\begin{equation}(\partial^2+2\lambda v^2)H(x)=0.
\label{oyster}
\end{equation}

The  effective action terms determining the $H$ decay amplitude at one-loop order in terms of the propagators described previously are determined from (\ref{kagemucha}). One term of the effective action is:
\begin{eqnarray}&&
S_{I}=-2i \lambda e^2 v\int d^4x\int d^4yH(x) G_{{\rm sc}}(x, y){\cal A}^\nu(y)
{\cal A}_\nu(y) G_{\rm {sc}}(y,x)
\nonumber\\&&
-8\lambda e^2v\int d^4x\int d^4y\int d^4zH(x)G_{{\rm sc} }(x, y){\cal A}^\nu(y)
D_\nu G_{{\rm sc}}(y, z)
{\cal A}^\lambda(z)
D_\lambda G_{{\rm sc}}(z, x)
\label{dulwich}
\end{eqnarray}
which is a seagull term and a derivative coupling term in the way familiar  from scalar quantum electrodynamics, with a Higgs boson insertion in one propagator. The remaining effective action terms are (\ref{arason})-(\ref{frederik}) listed in app. A. Remarkably, they can be reduced to a structure similar to (\ref{dulwich}), with both scalar and vector internal propagators, and in the latter case also with magnetic moment couplings. The reduction takes place by means of (\ref{ehuduanna}) and (\ref{Ward}).

In  (\ref{arason})     one isolates the following three expressions by  insertion of (\ref{supergill}):
\begin{eqnarray}&&
S_{II}'=-ie^2gM_W\int d^4yH(x)G_{{\rm vec}}\hspace{0.1 mm}^{\mu \lambda}(x, y){\cal A}^\nu (y){\cal A}_\nu (y)G_{{\rm vec},  \lambda \mu }(y, x)
\nonumber\\&&
+4e^2g M_W \int d^4x\int d^4y\int d^4zH(x)
G_{{\rm vec},  \mu}\hspace{0.1 mm}^{ \rho}(x, y)
\nonumber\\&&
{\cal A}^\nu (y)D_\nu G_{{\rm vec},  \rho}\hspace{0.1 mm}^{ \sigma}(y, z)
{\cal A}^{\lambda}(z)D_\lambda G_{{\rm vec},  \sigma}\hspace{0.1 mm}^{ \mu}(z, x),
\label{laramie}
\end{eqnarray}
which obviously is similar to (\ref{dulwich}),
\begin{eqnarray}&&
S_{III}'=4e^2 g M_W \int d^4x\int d^4y\int d^4zH(x)  
\nonumber\\&&
G_{{\rm vec},  \mu \lambda }(x, y){\cal F}  ^{\lambda \rho }(y )G_{{\rm vec},  \rho \sigma  }(y, z){\cal F} ^{\sigma \omega}  (z )
 G_{{\rm vec}, \omega}\hspace{0.1 mm}^\mu (z, x),
\label{appalachian}
\end{eqnarray}
with magnetic moment couplings, and:
\begin{eqnarray}&&
S_{IV}'=-4e^2 g M_W \int d^4x\int d^4y\int d^4zH(x)
\nonumber\\&&
(G_{{\rm vec},  \mu \rho}(x, y){\cal F}  ^{\rho\sigma }(y )G_{{\rm vec},  \sigma \omega }(y, z){\cal A}^\lambda(z)
D_\lambda G_{{\rm vec}, }\hspace{0.1 mm}^{\omega \mu} (z, x)
\nonumber\\&&
+G_{{\rm vec}, \mu \rho}(x, y){\cal A}^\nu(x)D_\nu G_{{\rm vec}, }\hspace{0.1 mm}^ {\rho}\hspace{0.1 mm}_{ \omega }(y, z) {\cal F} ^{\omega\epsilon}  (z ) G_{{\rm vec}, \epsilon}\hspace{0.1 mm}^\mu (z, x)).
\label{athapaskian}
\end{eqnarray}
with a derivative coupling at one vertex and a magnetic moment coupling at the other vertex.
Adding the rest of (\ref{arason}) to (\ref{ararat})-(\ref{frederik})  one obtains as shown in app. A:
\begin{eqnarray}&&
S_{V}'=ie ^2gM_W\int d^4x\int d^4yH(x)G_{{\rm sc}}(x, y)
{\cal A}^\nu  {\cal A}_\nu (y) G_{{\rm sc}} (y, x)
\nonumber\\&&
+4e^2gM_W\int d^4x\int d^4y\int d^4zH(x) G_{{\rm sc}}(x, y){\cal A}^\nu (y)
D _{\nu}
G_{{\rm sc}}(y, z)
{\cal A}^\lambda (z)D_{ \lambda } G_{{\rm sc}}(z, x)
\label{bedini}
\end{eqnarray}
with the same structure as (\ref{dulwich}) or (\ref{laramie}). 

\begin{figure}[ht]
\centering
\includegraphics[width=0.75\textwidth]{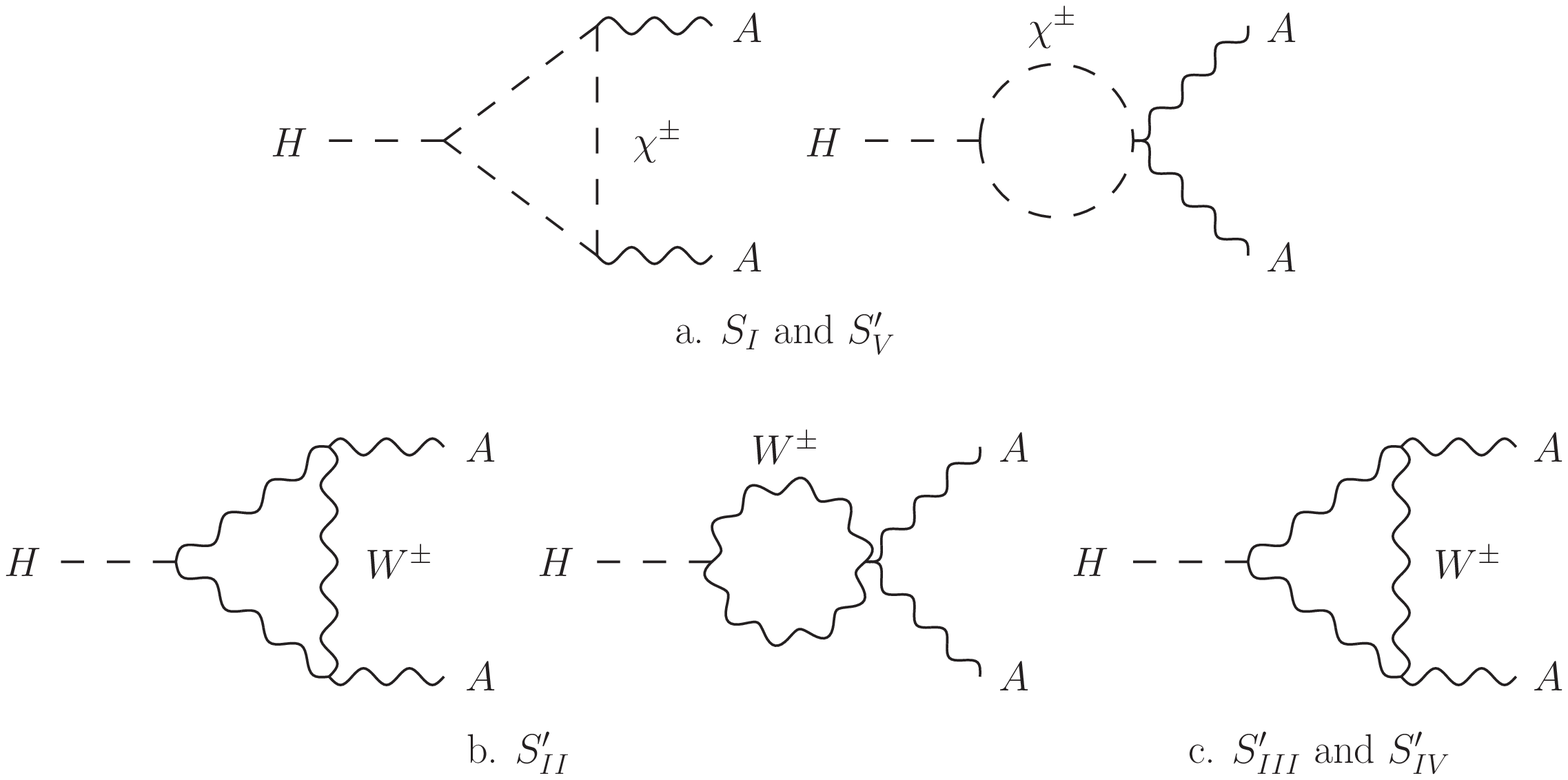}
\caption{Feynman diagram representation of the effective action in its final form.}
\label{fig:fig1}
\end{figure}

A Feynman diagram representation of $S_I$ and $S_{II}'-S_V'$ is given in Figure \ref{fig:fig1}.

\subsection{$H\rightarrow \gamma\gamma$  decay amplitude in a vanishing external field}

From (\ref{dulwich}) and (\ref{laramie})-(\ref{bedini})  the amplitude of the decay of a Higgs boson  to two photons is found.
Here and elsewhere in the paper the photon momenta and polarization vectors are denoted $k, \varepsilon_\mu(k)$ and $q, \varepsilon_\nu(q)$, with $k\cdot \epsilon(k)=q\cdot \epsilon(q)=0$. The evaluation is carried out by means of (\ref{ivantaurus}), (\ref{kenneweg}), (\ref{timian}) and (\ref{ginsberg}).

 In the limit where the background field vanishes  the contribution from  
 (\ref{laramie}) to the amplitude  is  in the proper-time representation:
\begin{eqnarray}&&
-8i  e^2gM_W\varepsilon^\mu(k)\varepsilon_\mu(q)
\int _0^\infty \tau d\tau \int _0^1d\alpha \int \frac {d^4r}{(2\pi)^4}e^{i\tau ((1-\alpha )r^2+\alpha (p-r)^2-M_W^2)}
\nonumber\\&&
+16 e^2gM_W\varepsilon^\mu(k)\varepsilon^\nu(q)
\int _0^\infty \tau^2d\tau e^{-i\tau M_W^2}
\int _0^1d\alpha d\beta d\gamma
 \delta (1-\alpha -\beta -\gamma)
\nonumber\\&&
\int \frac {d^4r}{(2\pi)^4}\bigg (r_\mu (r+k)_\nu e^{i\tau (\alpha r^2 +\beta (k+r)^2+\gamma (k+q+r)^2)}
+r_\nu (r+q)_\mu
 e^{i\tau (\alpha r^2 +\beta (q+r)^2+\gamma (k+q+r)^2)}\bigg)
\nonumber\\&&
\label{corydon}
\end{eqnarray}
where the integrations of the proper time $\tau$ are  carried out after the integrations of the momentum variable $r$; the momentum integrations  are convergent at nonvanishing values of the proper time. Here  a factor $(2\pi)^4\delta (p-k-q)$ is suppressed, with $p$ the Higgs boson momentum. After some manipulations one gets from (\ref{corydon}), using the mass-shell conditions as well as symmetric integration in four dimensions: 
\begin{eqnarray}&&
-8i  e^2gM_W\varepsilon^\mu(k)\varepsilon_\mu(q)
\int _0^\infty \tau d\tau \int _0^1d\alpha \int \frac {d^4r}{(2\pi)^4}e^{i\tau (r^2+\alpha(1-\alpha)M_H^2-M_W^2)}
\nonumber\\&&
+16e^2gM_W\varepsilon^\mu(k)\varepsilon_\mu(q)
\int _0^\infty \tau^2d\tau e^{-i\tau M_W^2}
\int _0^1d\alpha   \int _0^{1-\alpha} d\gamma
\int \frac {d^4r}{(2\pi)^4}e^{i\tau (r^2+\alpha \gamma M_H^2)}(\frac 12 r^2 -\alpha \gamma M_H^2  )
\nonumber\\&&
+32 e^2gM_W\varepsilon^\mu(k)\varepsilon^\nu(q) (q\cdot k \eta_{\mu \nu}-q_\mu k_\nu)
\int _0^\infty \tau^2d\tau e^{-i\tau M_W^2}
\int _0^1d\alpha \int _0^{1-\alpha} d\gamma \alpha \gamma 
\int \frac {d^4r}{(2\pi)^4}e^{i\tau (r^2+\alpha \gamma M_H^2)}.
\nonumber\\&&
\label{raketcykel}
\end{eqnarray}
In (\ref{raketcykel}) one uses:
\begin{equation}
\int \frac{d^4r}{(2 \pi)^4}  e^{i \tau r^2} 
=-\frac{i}{16\pi ^2\tau ^2}
\label{legolas}
\end{equation}
and also:
\begin{eqnarray}&&
\int _0^\infty \tau^2d\tau e^{-i\tau M_W^2}
\int _0^1d\alpha   \int _0^{1-\alpha} d\gamma
\int \frac {d^4r}{(2\pi)^4}e^{i\tau (r^2+\alpha \gamma M_H^2)}(\frac 12 r^2 -\alpha \gamma M_H^2  )
\nonumber\\&&
= \frac 12\frac{1}{16\pi^2}\int _0^\infty \frac{ d\tau }{\tau}e^{-i\tau M_W^2}
\int _0^1d\alpha   
e^{i\alpha (1-\alpha) \tau M_H^2}.
\label{thranduil}
\end{eqnarray}
Evaluating   (\ref{raketcykel}) by (\ref{legolas}) and (\ref{thranduil}) one finds that the  first two terms  cancel out, and (\ref{raketcykel})  reduces to:
\begin{eqnarray}&&
-\frac{2e^2gM_W}{\pi^2}\varepsilon^\mu(k)\varepsilon^\nu(q) (q\cdot k \eta_{\mu \nu}-q_\mu k_\nu) \int _0^1d\alpha \int _0^{1-\alpha} d\gamma
 \frac{\alpha \gamma}{M_W^2-\alpha \gamma M_H^2}
\nonumber\\&&
=\frac{ e^2gM_W}{\pi^2M_H^2}\varepsilon^\mu(k)\varepsilon^\nu(q) (q\cdot k \eta_{\mu \nu}-q_\mu k_\nu) (1-\frac{4M_W^2}{M_H^2}\arcsin^2(\frac{M_H}{2M_W})).
\label{hugodrax}
\end{eqnarray}
The total contribution to the amplitude from (\ref{dulwich}), (\ref{laramie}) and  (\ref{bedini})  is found from (\ref{hugodrax}) by the substitution:
\begin{equation}
4e^2gM_W\rightarrow 2\lambda e^2v+3e^2gM_W.
\label{hanebane}
\end{equation}

(\ref{appalachian})  in a vanishing external field contributes to the decay amplitude:
\begin{eqnarray}&&
-4 e^2gM_W   (k^\mu \varepsilon^\nu (k)-k^\nu \varepsilon^\mu (k))
(q_\mu \varepsilon_{\nu}(q)-q_\nu  \varepsilon_{\mu}(q))
\nonumber\\&&
\int _0^\infty \tau ^2d\tau e^{-i\tau M_W^2}\int _0^1 d\alpha d\beta d\gamma \delta(1-\alpha-\beta -\gamma)
\nonumber\\&&
 \int \frac {d^4r}{(2\pi)^4}\bigg ( e^{i\tau (\alpha r^2 +\beta (k+r)^2+\gamma (k+q+r)^2)}
+
 e^{i\tau (\alpha r^2 +\beta (q+r)^2+\gamma (k+q+r)^2)}\bigg)
 \nonumber\\&&
=\frac{2e^2gM_W}{\pi^2 M_H^2} \varepsilon^\mu(k)\varepsilon^\nu(q) (q\cdot k \eta_{\mu \nu}-q_\mu k_\nu)  \arcsin^2(\frac{M_H}{2M_W}).
 \label{ilderpeter}
\end{eqnarray}
(\ref{athapaskian}) is zero  in a vanishing external field.

The decay amplitude with vanishing external field  is the sum of (\ref{hugodrax}) (with the substitution (\ref{hanebane})) and   (\ref{ilderpeter}):
\begin{eqnarray}&&
\frac{e^2}{4\pi^2v}\varepsilon^\mu(k)\varepsilon^\nu(q) (q\cdot k \eta_{\mu \nu}-q_\mu k_\nu)
\nonumber\\&&
((1+\frac{6M_W^2}{M_H^2})(1-\frac{4M_W^2}{M_H^2}\arcsin^2(\frac{M_H}{2M_W}))+\frac{16M_W^2}{M_H^2}\arcsin^2(\frac{M_H}{2M_W}))
\label{eleanor}
\end{eqnarray}
which is the standard decay amplitude \cite{Ellis}-\cite{Marciano}. It is perhaps an interesting point that this result has been obtained by proper-time regularization instead of dimensional regularization; symmetrical integration in momentum space has been carried out in four dimensions and this is possible because momentum integrals are finite at  nonvanishing values of the proper time $\tau$.

 Carrying for the sake of argument  the integral in (\ref{hugodrax}) out in two space-time dimensions  one gets, disregarding the dimensional mismatch,  the result:
\begin{eqnarray}&&
-\frac{8 e^2gM_W}{\pi} \int _0^1d\alpha \int _0^{1-\alpha} d\gamma
 \frac{\alpha \gamma}{(M_W^2-\alpha \gamma M_H^2)^2}
\nonumber\\&&
=-\frac{8 e^2 gM_W}{\pi M_H^4}
(\frac{M_H}{\sqrt{M_W^2-\frac 14M_H^2}}\arcsin (\frac{M_H}{2M_W})-2\arcsin ^2(\frac{M_H}{2M_W}))
\label{rimelig}
\end{eqnarray}
This is singular at $\frac{M_H}{M_W}=2$; the singularity arises from $\alpha \simeq \gamma \simeq \frac 12$ where the denominator of the integrand is very small at this value of the mass ratio. This argument gives a heuristic indication of the way in which the square-root singularity of (\ref{manderley}) arises, since the quasi-tachyonic field component decreases the vector boson mass according to:
\begin{equation}
M_W^2\rightarrow M_W^2-eB.
\label{rebecca}
\end{equation}
The complete determination of the singularity takes place in sec. 4.

\section{$H\rightarrow \gamma\gamma$  decay amplitude in a non-vanishing  homogeneous field}

The $H\rightarrow \gamma \gamma$ amplitude in a non-vanishing  homogeneous electromagnetic field is found from (\ref{dulwich})-(\ref{bedini}) by the method of Schwinger \cite{Schwinger},\cite{Adler}, \cite{Tsai}. Details on formal tools are relegated to App.B.

The contribution from  the first term of (\ref{dulwich})  to the decay amplitude is by (\ref{dynamide}):
\begin{eqnarray}&&
-4i \lambda e^2v \varepsilon^\mu(k)\varepsilon_\mu(q)
\int d^4x e^{ipx}\int _0^\infty \tau d\tau e^{-i\tau M_W^2}\int _0^1d\alpha <x\mid  e^{i(1-\alpha) \tau\Pi^2}e^{-i(k+q)X}e^{i\alpha \tau\Pi^2}\mid x>
\nonumber\\&&
=-4i \lambda e^2v\varepsilon^\mu(k)\varepsilon_\mu(q)
\int d^4x e^{ipx}\int _0^\infty \tau d\tau e^{-i\tau M_W^2}\int _0^1d\alpha <x, \tau \mid  e^{-i(k+q)X(\alpha \tau)}\mid x, 0>.
\label{queensland}
\end{eqnarray}
 Here one uses (\ref{boong}), as well as the eigenvalue  equation (\ref{lassalle}),
to  get the following value of (\ref{queensland}) with a factor $(2\pi)^4\delta(p-k-q)$ suppressed (cf. \cite{Tsai}):
\begin{eqnarray} 
-4i \lambda e^2v \varepsilon^\mu(k)\varepsilon_\mu(q)
\int _0^\infty \tau d\tau e^{-i\tau M_W^2}<x, \tau\mid x, 0>\int _0^1d\alpha e^{\delta _1(\alpha,  k+q)}.
\label{plausibel}
\end{eqnarray}

Next the contribution of  the second term of (\ref{dulwich})  to the decay amplitude is evaluated. It has the proper-time representation:
\begin{eqnarray}&&
8\lambda e^2v\varepsilon^\mu(k)\varepsilon^\nu(q)\int d^4x e^{ipx}
\int _0^\infty \tau^2d\tau e^{-i\tau M_W^2}\int _0^1d\alpha d\beta d\gamma
 \delta (1-\alpha -\beta -\gamma)
 \nonumber\\&&
(<x\mid e^{i\alpha \tau\Pi^2}e^{-ik\cdot X} \Pi_\mu e^{i\beta \tau\Pi^2} e^{-iq\cdot X}\Pi_\nu e^{i\gamma \tau\Pi^2}\mid x>+(\mu \leftrightarrow \nu, k \leftrightarrow q))
\nonumber\\&&
=
8\lambda e^2v\varepsilon^\mu(k)\varepsilon^\nu(q)
\int d^4x e^{ipx}\int _0^\infty \tau^2d\tau e^{-i\tau M_W^2}\int _0^1d\alpha d\beta d\gamma
 \delta (1-\alpha -\beta -\gamma)
 \nonumber\\&&
(<x, \tau \mid \Pi_\mu ((1-\alpha )\tau)  e^{-ik\cdot X((1-\alpha )\tau} e^{-iq\cdot X(\gamma\tau )}\Pi_\nu(\gamma\tau) \mid x, 0>+(\mu \leftrightarrow \nu, k \leftrightarrow q))
\label{foxtrot}
\end{eqnarray}
by (\ref{trudeau}), and using here  (\ref{classicgrandcru}) and (\ref{mahelia}) as well as the procedure used above to obtain (\ref{plausibel}) one finds:
\begin{eqnarray}&&
8\lambda e^2v\varepsilon^\mu(k)\varepsilon^\nu(q)
\int d^4x e^{ipx}\int _0^\infty \tau^2d\tau e^{-i\tau M_W^2}\int _0^1d\alpha d\beta d\gamma
 \delta (1-\alpha -\beta -\gamma)
\nonumber\\&&
(e^{\delta_2(k, q)}
<x, \tau \mid  \Pi_\mu  ((1-\alpha )\tau)  e^{-iQ\cdot X(\tau )}e^{-i(k+q-Q)\cdot X(0)}\Pi_\nu(\gamma\tau) \mid x, 0>+(\mu \leftrightarrow \nu, k \leftrightarrow q))
\nonumber\\&&
=8\lambda e^2v\varepsilon^\mu(k)\varepsilon^\nu(q)(2\pi)^4\delta(p-k-q)
\int _0^\infty \tau^2d\tau e^{-i\tau M_W^2}\int _0^1d\alpha d\beta d\gamma
 \delta (1-\alpha -\beta -\gamma)  
\nonumber\\&&
(e^{\delta_2(k, q)}<x, \tau \mid (\Pi((1-\alpha )\tau)-e^{2\alpha\tau e{\bf F}}Q)_\mu
 (\Pi(\gamma \tau)+(e^{-2\gamma \tau e{\bf F}}(k+q-Q))_\nu \mid x, 0>
\nonumber\\&&
+(\mu \leftrightarrow \nu, k \leftrightarrow q))
\label{thorbellinge}
\end{eqnarray}
 and in the last step (\ref{trudeau}) was used again.

The evaluation of (\ref{thorbellinge}) is carried out by (\ref{trudeau}) and (\ref{borkrigel}). Only terms with two or no $\Pi$ operators  give a nonvanishing contribution. With no $\Pi$ operators one gets the following contribution from (\ref{thorbellinge}):
\begin{eqnarray}&&
-8\lambda e^2v\varepsilon^\mu(k)\varepsilon^\nu(q)
\int _0^\infty \tau^2d\tau e^{-i\tau M_W^2} <x, \tau \mid x,0>\int _0^1d\alpha d\beta d\gamma
 \delta (1-\alpha -\beta -\gamma) 
\nonumber\\&&
(e^{\delta_2(k, q)}
(e^{2\alpha \tau e{\bf F}}Q)_\mu(e^{-2\gamma \tau e{\bf F}}(k+q-Q))_\nu+(\mu \leftrightarrow \nu, k \leftrightarrow q)).
\label{frodobaggins}
\end{eqnarray}
The term of (\ref{thorbellinge}) with two $\Pi$ operators contributes: 
\begin{eqnarray}&&
8i\lambda e^2v\varepsilon^\mu(k)\varepsilon^\nu(q)
\int _0^\infty \tau^2d\tau e^{-i\tau M_W^2} <x, \tau \mid x,0>\int _0^1d\alpha d\beta d\gamma
 \delta (1-\alpha -\beta -\gamma) \nonumber\\&&
(e^{\delta_2(k, q)} 
 (e^{-2\beta \tau e{\bf F}}{\bf D}^{-1}(\tau))_{\mu \nu}+(\mu \leftrightarrow \nu, k \leftrightarrow q)).
 \label{lashmar}
 \end{eqnarray}
In both (\ref{frodobaggins}) and (\ref{lashmar}) a factor $(2\pi)^4\delta(p-k-q)$ was left out.  The sum   of (\ref{plausibel}),  (\ref{frodobaggins}) and (\ref{lashmar}) is invariant under gauge transformations of the polarization vectors. This  follows from the general proof in  (\ref{skarphedin}) (app. C) but can  be proven directly  also.

(\ref{plausibel}) and (\ref{lashmar}) are both ultraviolet divergent  and can be rearranged in two convergent expressions:
\begin{eqnarray}&&
8i\lambda e^2v\varepsilon^\mu(k)\varepsilon^\nu(q)
\int _0^\infty \tau^2d\tau e^{-i\tau M_W^2} <x, \tau \mid x,0>\int _0^1d\alpha d\beta d\gamma
 \delta (1-\alpha -\beta -\gamma) \nonumber\\&&
(e^{\delta_2(k, q)} 
 (e^{-2\beta \tau e{\bf F}}{\bf D}^{-1}(\tau)-\frac{1}{2\tau}{\bf 1})_{\mu \nu}+(\mu \leftrightarrow \nu, k \leftrightarrow q))
 \label{marlash}
 \end{eqnarray}
and:
\begin{eqnarray}&&
4i\lambda e^2v\varepsilon^\mu(k)\varepsilon_\mu(q)
\int _0^\infty \tau d\tau e^{-i\tau M_W^2} <x, \tau \mid x,0>\int _0^1d\alpha d\beta d\gamma
 \delta (1-\alpha -\beta -\gamma) \nonumber\\&&
((e^{\delta_2(k, q)} -e^{\delta_2(k, q)} \mid _{\gamma =1-\alpha})
 +(k \leftrightarrow q)).
 \label{mehemet}
 \end{eqnarray}

 The contribution of (\ref{laramie})  to the amplitude  is:
\begin{eqnarray}&&
-2ie^2g M_W \varepsilon^\mu(k)\varepsilon_\mu(q) 
\int d^4x e^{ipx}\int _0^\infty \tau d\tau e^{-i\tau M_W^2}{\rm tr}(e^{-2 \tau  e {\bf  F}})
\nonumber\\&&
\int _0^1d\alpha <x\mid  e^{i(1-\alpha) \tau\Pi^2}e^{i(k+q)X}e^{i\alpha \tau\Pi^2}\mid x>
\nonumber\\&&
+4e^2gM_W \varepsilon^\mu(k)\varepsilon^\nu(q) 
\int d^4xe^{ipx}\int _0^\infty \tau^2d\tau e^{-i\tau M_W^2}\int _0^1d\alpha d\beta d\gamma
 \delta (1-\alpha -\beta -\gamma)
\nonumber\\&&
{\rm tr}(e^{-2 \tau  e {\bf  F}})(<x\mid e^{i\alpha \tau\Pi^2}\Pi _\mu e^{ik\cdot X}e^{i\beta \tau\Pi^2}\Pi _\nu e^{i\gamma \tau\Pi^2}\mid x>+(\mu \leftrightarrow \nu, k \leftrightarrow q))
\label{kirsewetter}
\end{eqnarray}
and is thus determined from  (\ref{frodobaggins}), (\ref{marlash}) and (\ref{mehemet}) by the substitution $8\lambda e^2v\rightarrow 4e^2gM_W$ and insertion of a factor ${\rm tr}(e^{-2 \tau  e {\bf  F}})$ in the integral. Also from    (\ref{bedini}) one gets three terms similar to   (\ref{frodobaggins}), (\ref{marlash}) and (\ref{mehemet})    by the substitution $8\lambda e^2v\rightarrow -4e^2gM_W.$

For the considerations on a pure magnetic field in the following section it is convenient to isolate in the contribution to the amplitude from (\ref{kirsewetter})
 the following three terms:
\begin{eqnarray}&&
-4 e^2gM_W\varepsilon^\mu(k)\varepsilon^\nu(q) 
\int _0^\infty \tau ^2d\tau e^{-i\tau M_W^2}({\rm tr}(e^{-2 \tau  e {\bf  F}}) -4)<x, \tau \mid x,0> \delta (1-\alpha -\beta -\gamma)   
 \nonumber\\&&
 (e^{\delta_2(k, q)}
(e^{2\alpha \tau e{\bf F}}Q)_\mu(e^{-2\gamma \tau e{\bf F}}(k+q-Q))_\nu+(\mu \leftrightarrow \nu, k \leftrightarrow q))
\label{hickory}
\end{eqnarray}
and also:
\begin{eqnarray}&&
4i  e^2gM_W\varepsilon^\mu(k)\varepsilon^\nu(q)
\int _0^\infty \tau ^2d\tau e^{-i\tau M_W^2}({\rm tr}(e^{-2 \tau  e {\bf  F}}) -4)<x, \tau \mid x,0>
\int _0^1d\alpha d\beta d\gamma
 \delta (1-\alpha -\beta -\gamma) 
 \nonumber\\&&
 (e^{\delta_2(k, q)} 
 (e^{-2\beta \tau e{\bf F}}{\bf D}^{-1}(\tau)-\frac{1}{2\tau}{\bf 1})_{\mu \nu}+(\mu \leftrightarrow \nu, k \leftrightarrow q))
   \label{niagara}
\end{eqnarray}
and:
\begin{eqnarray}&&
2i e^2gM_W\varepsilon^\mu(k)\varepsilon_\mu(q) 
\int _0^\infty \tau d\tau e^{-i\tau M_W^2}({\rm tr}(e^{-2 \tau  e {\bf  F}})-4) <x, \tau \mid x,0>
\int _0^1d\alpha  d\beta d\gamma
 \delta (1-\alpha -\beta -\gamma) 
 \nonumber\\&&
( (e^{\delta_2(k, q)} -e^{\delta_2(k, q)}\mid _{\gamma=1-\alpha})+(k\leftrightarrow q))
 \label{donaulloyd}
\end{eqnarray}
and to further isolate in (\ref{hickory}) and (\ref{donaulloyd}):
\begin{eqnarray}&&
-4 e^2gM_W\varepsilon^\mu(k)\varepsilon^\nu(q) 
\int _0^\infty \tau ^2d\tau e^{-i\tau M_W^2}({\rm tr}(e^{-2 \tau  e {\bf  F}}) -4)<x, \tau \mid x,0>\int _0^1d\alpha d\beta d\gamma \alpha \gamma\delta (1-\alpha -\beta -\gamma)   
 \nonumber\\&&
 (e^{\delta_2(k, q)}+e^{\delta_2(q, k)})
(q_\mu k_\nu-\eta _{\mu \nu}q\cdot k).
\label{ambeno}
\end{eqnarray}
Here was used:
\begin{eqnarray}&&
\int _0^1d\alpha d\beta d\gamma \delta (1-\alpha-\beta-\gamma)(e^{\delta_2(k,q)}-e^{\delta_2(k,q)}\mid _{\gamma=1-\alpha})
\nonumber\\&&
=-\int _0^1d\alpha d\beta d\gamma \delta (1-\alpha-\beta-\gamma)\frac 12 (\alpha\frac{\partial}{\partial \alpha}+\gamma 
 \frac{\partial}{\partial \gamma}) e^{\delta_2(k,q)}
\label{hornhyl}
\end{eqnarray}
and also (\ref{octopussy}).
The remaining amplitude  terms from (\ref{laramie})    and  (\ref{bedini}) are obtained from (\ref{frodobaggins}),  (\ref{marlash}) and (\ref{mehemet}) by the substitution $2\lambda e^2v\rightarrow 3e^2gM_W$.

From  (\ref{appalachian}) one gets:
 \begin{eqnarray}&&
4 e^2gM_W \varepsilon^\mu(k)\varepsilon^\nu(q)
\int d^4x e^{ipx}\int _0^\infty \tau^2d\tau e^{-i\tau M_W^2}\int _0^1d\alpha d\beta d\gamma
 \delta (1-\alpha -\beta -\gamma) 
\nonumber\\&&
\bigg((e^{-2(\alpha+\gamma) \tau e {\bf F}})_{\epsilon \rho }(\delta ^{\rho}\hspace{0.1 mm}_\mu k^\sigma - \delta ^{\sigma}\hspace{0.1 mm}_\mu k^\rho) 
 (e^{-2\beta \tau e{\bf F}})_{\sigma \omega} (\delta ^{\omega}\hspace{0.1 mm}_\nu q^\epsilon - \delta ^{\epsilon}\hspace{0.1 mm}_\nu q^\omega) 
\nonumber\\&&
 <x\mid e^{i\alpha \tau\Pi^2}e^{-ik\cdot X}  e^{i\beta \tau\Pi^2} e^{-iq\cdot X}\ e^{i\gamma \tau\Pi^2}\mid x>
+(\mu \leftrightarrow \nu, k \leftrightarrow q)\bigg)
\label{ordovician}
\end{eqnarray}
which after similar manipulations as were used to obtain (\ref{plausibel}) gives the amplitude term: 
\begin{eqnarray}&&
 4 e^2gM_W\varepsilon^\mu(k)\varepsilon^\nu(q)
\int _0^\infty \tau^2d\tau e^{-i\tau M_W^2}<x, \tau \mid x,0>\int _0^1d\alpha d\beta d\gamma
 \delta (1-\alpha -\beta -\gamma)
\nonumber\\&&
\bigg(e^{\delta_2(k, q)}
 (e^{-2(\alpha+\gamma) \tau e{\bf F}})_{\epsilon \rho }(\delta ^{\rho}\hspace{0.1 mm}_\mu k^\sigma - \delta ^{\sigma}\hspace{0.1 mm}_\mu k^\rho) 
 (e^{-2\beta \tau e{\bf F}})_{\sigma \omega} (\delta ^{\omega}\hspace{0.1 mm}_\nu q^\epsilon - \delta ^{\epsilon}\hspace{0.1 mm}_\nu q^\omega)+(\mu \leftrightarrow \nu, k \leftrightarrow q)\bigg).
\nonumber\\&&
  \label{ledreborg}
 \end{eqnarray}
Also (\ref{athapaskian}) yields:
\begin{eqnarray}&&
-4e^2 g M_W\varepsilon^\mu(k)\varepsilon^\nu(q)
\int d^4x e^{ipx}\int _0^\infty \tau^2d\tau e^{-i\tau M_W^2}
\int _0^1d\alpha d\beta d\gamma
 \delta (1-\alpha -\beta -\gamma)
 \nonumber\\&&
\bigg ( ((e^{-2\tau e{\bf F}})_{\sigma \rho}(\delta ^{\rho}\hspace{0.1 mm}_\mu k^\sigma - \delta ^{\sigma}\hspace{0.1 mm}_\mu k^\rho)<x\mid e^{i\alpha \tau \Pi^2} e^{-ik\cdot X} e^{i\beta \tau \Pi^2}
 e^{-iq\cdot X}\Pi_\nu e^{i\gamma \tau \Pi^2}\mid x>
 \nonumber\\&&
 +(e^{-2\tau e{\bf F}})_{\epsilon \omega}(\delta ^{\omega}\hspace{0.1 mm}_\nu q^\epsilon - \delta ^{\epsilon}\hspace{0.1 mm}_\nu q^\omega) <x\mid e^{i\alpha \tau \Pi^2} e^{ik\cdot X}\Pi_\mu e^{i\beta \tau \Pi^2}
 e^{iq\cdot X} e^{i\gamma \tau \Pi^2}\mid x>)
\nonumber\\&&
+(\mu \leftrightarrow \nu, k \leftrightarrow q)\bigg)
\label{wullenwever}
\end{eqnarray}
that is evaluated in a similar way, contributing to the amplitude:
 \begin{eqnarray}&&
  -4e^2 g M_W \varepsilon^\mu(k)\varepsilon^\nu(q)
\int _0^\infty \tau^2d\tau e^{-i\tau M_W^2}<x, \tau \mid x, 0>
\int _0^1d\alpha d\beta d\gamma
 \delta (1-\alpha -\beta -\gamma)
 \nonumber\\&&
\bigg(e^{\delta _2(k, q)} ((e^{-2\tau e{\bf F}})_{\sigma \rho}(\delta ^{\rho}\hspace{0.1 mm}_\mu k^\sigma - \delta ^{\sigma}\hspace{0.1 mm}_\mu k^\rho)(e^{-2\gamma \tau e{\bf F}}(k+q-Q))_\nu
 \nonumber\\&&
 -(e^{-2\tau e{\bf F}})_{\epsilon \omega}(\delta ^{\omega}\hspace{0.1 mm}_\nu q^\epsilon - \delta ^{\epsilon}\hspace{0.1 mm}_\nu q^\omega) (e^{2\alpha\tau e{\bf F}}Q)_\mu)
+(\mu \leftrightarrow \nu, k \leftrightarrow q)\bigg).
 \label{stribenlanz}
\end{eqnarray}

\section{$H\rightarrow \gamma\gamma$ decay amplitude in a pure magnetic field}

The $H\rightarrow \gamma\gamma$ decay amplitude is considered in a pure homogeneous magnetic field $B$ directed along the positive $1$-axis,  with $k^2=q^2=0, 2k\cdot q=M_H^2$.

In this case (\ref{ambeno}) is, in the special case where the  photons are emitted along   the magnetic field lines, using also  (\ref{debeers}) combined with (\ref{minimal}) as well as (\ref{macaroni}) and  (\ref{benazir}):
\begin{eqnarray}&&
-\frac{2i}{\pi^2}e^2gM_W \varepsilon^\mu(k)\varepsilon^\nu(q)(q_\mu k_\nu-\eta _{\mu \nu}q\cdot k)
\nonumber\\&&
\int _0^\infty d\tau e^{-i\tau M_W^2}\tau eB\sin(\tau eB) \int _0^1d\alpha d\beta d\gamma
\alpha \gamma \delta (1-\alpha -\beta -\gamma) e^{i\alpha \gamma \tau M_H^2}
\nonumber\\&&
=\frac{2}{\pi^2}e^2gM_W\frac{eB}{M_H^2}\varepsilon^\mu(k)\varepsilon^\nu(q)(q_\mu k_\nu-\eta _{\mu \nu}q\cdot k)
\nonumber\\&&
(-\frac{1}{M_H^2}(\arcsin^2(\frac{M_H}{2\sqrt{M_W^2-eB}})-\arcsin^2(\frac{M_H}{2\sqrt{M_W^2+eB}}))
\nonumber\\&&
+\frac{1}{2M_H}
(\frac{1}{\sqrt{M_W^2-eB-\frac{1}{4}M_H^2}}\arcsin(\frac{M_H}{2\sqrt{M_W^2-eB}}))
\nonumber\\&&
-\frac{1}{\sqrt{M_W^2+eB-\frac{1}{4}M_H^2}}\arcsin(\frac{M_H}{2\sqrt{M_W^2+eB}}))).
\label{beesiana}
\end{eqnarray}
(\ref{beesiana})  is divergent at $eB=M_W^2-\frac{1}{4}M_H^2$. This divergence can be attributed to the quasi-unstable mode of the $W^\pm$ field that decreases the effective  mass of a $W^\pm$ field component, combined with the fact that the magnetic field in a sense makes the theory two-dimensional since charged field modes only propagate along the field lines. This can also be seen from (\ref{rimelig}), which shows that one finds results similar to (\ref{beesiana})   redoing the calculation  of  the integrals determining the amplitude in a vanishing external field in sec.  2.4 in two  instead of four dimensions.

In the limit where the photon momenta vanish one may also obtain the amplitude from the  Heisenberg-Euler effective action. Having vanishing photon momenta one must  let the Higgs boson mass go to zero as well. (\ref{beesiana}) then becomes:
\begin{eqnarray}&&
-\frac{i}{12\pi^2}e^2gM_W \varepsilon^\mu(k)\varepsilon^\nu(q)(q_\mu k_\nu-\eta _{\mu \nu}q\cdot k)
\int _0^\infty d\tau e^{-i\tau M_W^2}\tau eB\sin(\tau eB) 
\nonumber\\&&
\simeq \frac{1}{24\pi^2} e^3gM_WB \varepsilon^\mu(k)\varepsilon^\nu(q)(q_\mu k_\nu-\eta _{\mu \nu}q\cdot k)(\frac{1}{(M_W^2-eB)^2}-\frac{1}{(M_W^2+eB)^2})
\label{maximilian}
\end{eqnarray}
and the square-root singularity is not visible in this limit.

The divergence arises at $\alpha\simeq \gamma \simeq \frac 12$ in which case the phase factor involving $\tau$ is constant in part of (\ref{beesiana}) and the $\tau$-integration diverges. That (\ref{beesiana}) is singular in this limit can also be seen directly by restricting both the Feynman parameters $\alpha $ and $\gamma$ in (\ref{beesiana}) to a narrow interval around $\frac 12$, in which case it is evaluated by  the following  calculation, with $0<\delta, \epsilon <<1$ (cf. \cite{Grisaru}):
\begin{eqnarray}&&
 \frac{e^3gM_WB}{4\pi^2}\varepsilon^\mu(k)\varepsilon^\nu(q)(q_\mu k_\nu-\eta _{\mu \nu}q\cdot k)
\int _{\frac 12-\epsilon}^{\frac 12+\epsilon} d\alpha  \int_{1-\alpha-\delta}^{1-\alpha} d\gamma \frac{1}{(M_W^2-eB-\alpha \gamma M_H^2)^2}
\nonumber\\&&
\simeq \frac{e^3gM_WB}{\pi^2M_H^3\sqrt{M_W^2-eB-\frac 14 M_H^2}}
\varepsilon^\mu(k)\varepsilon^\nu(q)(q_\mu k_\nu-\eta _{\mu \nu}q\cdot k)\arctan (\epsilon\frac{M_H}{\sqrt{M_W^2-eB-\frac 14M_H^2}})
\nonumber\\&&
\simeq \frac{e^3gM_WB}{2\pi M_H^3\sqrt{M_W^2-eB-\frac 14 M_H^2}}
\varepsilon^\mu(k)\varepsilon^\nu(q)(q_\mu k_\nu-\eta _{\mu \nu}q\cdot k)
\label{jemogfixiana}
\end{eqnarray}
where in the last step the $\arctan$ has been replaced by $\frac \pi 2$, which is valid with $\epsilon \neq 0$ kept fixed for $M_W^2-eB-\frac{1}{4}M_H^2\rightarrow 0,$
 and  (\ref{jemogfixiana}) agrees with (\ref{beesiana}) in this limit.
Here the contribution from the lower limit of the $\gamma$-integration was disregarded; it is finite at $M_W^2-eB-\frac{1}{4}M_H^2= 0$ for $\delta, \epsilon \neq 0$.

In  (\ref{jemogfixiana}) one can interchange the Feynman parameter integrations, observing that $\frac 12-\epsilon<\alpha <\frac 12 +\epsilon,  1-\alpha-\delta<\gamma <1-\alpha$ is equivalent to $\frac 12-\epsilon-\delta<\gamma <\frac 12+\epsilon, 1-\gamma - \delta <\alpha <1-\gamma$.

The singularity of  (\ref{ambeno}) is next determined  also with nonvanishing momentum components $\vec{k}_{\perp}, \vec{q}_{\perp}$ perpendicular to the magnetic field lines. The singularity  arises for  $\tau \rightarrow -i\infty, \alpha \simeq \gamma \simeq \frac 12$.  In this limit the quantity $\delta_2(k, q)$ is given by  (\ref{rottingham}) which  is nonlinear in the Feynman parameters $\alpha$ and $\gamma$, and the calculation is therefore more complicated than (\ref{jemogfixiana}). Approximating  $\delta_2(k, q)$  by the following expression:
\begin{eqnarray}&&
\delta_{2, {\rm app}}(k, q)=i\tau (\alpha \gamma (M_H^2+(\vec{q}_{\perp}+\vec{k}_{\perp})^2)+\frac 12(1-\alpha-\gamma)( \vec{k}_{\perp}^2+  \vec{q}_\perp^2))-\frac{1}{2eB}(\vec{k}_{\perp}+\vec{q}_{\perp})^2
\nonumber\\&&
-\frac{e^{-i\theta}}{2eB}(e^{-2i(1-\alpha-\gamma)\tau eB}-1)\mid \vec{q}_\perp\mid \mid \vec{k}_\perp\mid
\label{sangrail}
\end{eqnarray}
 with $\theta $ the angle between $\vec{k}_{\perp}$ and $ \vec{q}_{\perp}$ as defined in (\ref{lullubum}), one gets instead of (\ref{jemogfixiana}):
\begin{eqnarray}&&
-\frac{e^3gM_WB}{8\pi^2}\varepsilon^\mu(k)\varepsilon^\nu(q)(q_\mu k_\nu-\eta _{\mu \nu}q\cdot k)
\nonumber\\&&
\int _0^\infty \tau d\tau\int _{\frac 12-\epsilon}^{\frac 12+\epsilon} d\alpha  \int_{1-\alpha-\delta}^{1-\alpha}d\gamma e^{-i\tau(M_W^2-eB)}(e^{\delta_{2, {\rm app}}(k, q)}+(k\rightarrow q))
\nonumber\\&&
=\frac{e^2gM_W}{8\pi^2}\varepsilon^\mu(k)\varepsilon^\nu(q)(q_\mu k_\nu-\eta _{\mu \nu}q\cdot k)
\exp(-\frac{(\vec{k}_\perp+\vec{q}_\perp)^2}{2eB})
\nonumber\\&&
\int _{\frac 12-\epsilon}^{\frac 12+\epsilon} d\alpha  \int_{1-\alpha-\delta}^{1-\alpha}d\gamma
(\exp(\frac{e^{-i\theta}}{2eB}\mid \vec{q}_\perp\mid \mid \vec{k}_\perp\mid)\sum _{n=0}^\infty \frac{1}{n!}(-\frac{e^{-i\theta}\mid \vec{q}_\perp\mid \mid \vec{k}_\perp\mid}{2eB})^n
\nonumber\\&&
\frac{eB}{(M_W^2-eB-\alpha \gamma (M_H^2+(\vec{q}_{\perp}+\vec{k}_{\perp})^2)-(1-\alpha-\gamma)(\frac 12( \vec{k}_{\perp}^2+  \vec{q}_\perp^2)-2neB))^2}
\nonumber\\&&
+(\theta \rightarrow -\theta)).
\label{wolfgang}
\end{eqnarray}
The power series expansion has been carried out in order to make the $\tau$-integration possible. Next also the Feynman parameter integrations are carried out as in (\ref{jemogfixiana}):
\begin{eqnarray}&&
 \frac{e^2gM_W}{2\pi^2}\varepsilon^\mu(k)\varepsilon^\nu(q)(q_\mu k_\nu-\eta _{\mu \nu}q\cdot k)
\exp(-\frac{(\vec{k}_\perp+\vec{q}_\perp)^2}{2eB})
\nonumber\\&&
\int _{\frac 12}^{\frac 12+\epsilon} d\alpha \frac{eB}{M_W^2-eB-\frac{1}{4}(M_H^2+(\vec{k}_\perp+\vec{q}_\perp)^2)+(\alpha-\frac 12)^2  (M_H^2+(\vec{k}_\perp+\vec{q}_\perp)^2)}
\nonumber\\&&
(F(0, \theta)+F(0, -\theta))
\nonumber\\&&
\simeq \frac{e^2gM_W}{4 \pi}\varepsilon^\mu(k)\varepsilon^\nu(q)(q_\mu k_\nu-\eta _{\mu \nu}q\cdot k)
\exp(-\frac{(\vec{k}_\perp+\vec{q}_\perp)^2}{2eB})
\nonumber\\&&
\frac{eB}{\sqrt{(M_H^2+(\vec{k}_\perp+\vec{q}_\perp)^2)(M_W^2-eB-\frac 14 (M_H^2+(\vec{k}_{\perp}+\vec{q}_{\perp})^2))}}
(F(0, \theta)+F(0, -\theta))
\label{zinfandel}
\end{eqnarray}
 with the definition:
\begin{equation}
F(j, \theta)=\exp(\frac{e^{-i\theta}}{2eB}\mid \vec{q}_\perp\mid \mid \vec{k}_\perp\mid)
\sum _{n=0}^\infty \frac{1}{n!}(-\frac{e^{-i\theta}\mid \vec{q}_\perp\mid \mid \vec{k}_\perp\mid}{2eB})^n\frac{1}{M_H^2+2\vec{q}_{\perp} \cdot \vec{k}_{\perp}+4(n+j)eB}.
\label{armiger}
\end{equation}
(\ref{zinfandel}) is singular at $M_W^2-eB-\frac{1}{4}(M_H^2+(\vec{k}_\perp+\vec{q}_\perp)^2)\simeq 0$, with   
\begin{equation}
M_H^2+(\vec{k}_\perp+\vec{q}_\perp)^2=p_{0}^2-p_{1}^2,
\label{oxalsyre}
\end{equation}
where $p_{0}$ is the energy and $p_{ 1}$ the momentum along the magnetic field of the Higgs boson. One also notices the presence of an exponential damping factor $\exp(-\frac{(\vec{k}_\perp+\vec{q}_\perp)^2}{2eB})$.

Substituting in (\ref{ambeno}) the whole expression $\delta_2(k, q)$ as given by (\ref{rottingham}) one gets in addition to  (\ref{wolfgang}):
\begin{eqnarray}&&
\frac{e^2gM_W}{4\pi^2}\varepsilon^\mu(k)\varepsilon^\nu(q)(q_\mu k_\nu-\eta _{\mu \nu}q\cdot k)
\exp(-\frac{(\vec{k}_\perp+\vec{q}_\perp)^2}{2eB})
\nonumber\\&&\int _{\frac 12-\epsilon}^{\frac 12+\epsilon} d\alpha  \int_{1-\alpha-\delta}^{1-\alpha}d\gamma (1-\alpha-\gamma)((\alpha-\frac 12)\vec{k}_{\perp}^2+(\gamma-\frac 12)\vec{q}_{\perp}^2)
\nonumber\\&&
(\exp(\frac{e^{-i\theta}}{2eB}\mid \vec{q}_\perp\mid \mid \vec{k}_\perp\mid)\sum _{n=0}^\infty \frac{1}{n!}(-\frac{e^{-i\theta}\mid \vec{q}_\perp\mid \mid \vec{k}_\perp\mid}{2eB})^n
\nonumber\\&&
\int_0^1dt\frac{eB}{(M_W^2-eB-\alpha \gamma (M_H^2+(\vec{q}_{\perp}+\vec{k}_{\perp})^2)+(1-\alpha-\gamma)(2neB
-t((\alpha-\frac 12) \vec{k}_{\perp}^2+  (\gamma-\frac 12)\vec{q}_\perp^2)))^3}
\nonumber\\&&
+(\theta \rightarrow -\theta)).
\label{hornblower}
\end{eqnarray}
(\ref{hornblower}) is finite at $M_W^2-eB-\frac{1}{4}(M_H^2+(\vec{k}_\perp+\vec{q}_\perp)^2)\simeq 0$ as seen by changing to polar coordinates in the Feynman parameter space with origin at $\alpha=\gamma =\frac 12$. Consequently the singularity of (\ref{zinfandel}) is not modified by (\ref{hornblower}).

It has been demonstrated that  the singular behavior found in (\ref{beesiana}) or (\ref{jemogfixiana}) persists when the two photons produced in the decay also have momentum components orthogonal to the magnetic field, with the square root denominator modified as seen from  (\ref{zinfandel}) and with an exponential damping factor. For the sake of completeness it is now shown that the singularity, as well as the exponential damping factor found in (\ref{zinfandel}), occur in the complete expressions (\ref{hickory}), (\ref{niagara}) and (\ref{donaulloyd}) as well as in (\ref{ledreborg}) and (\ref{stribenlanz}).

The singular part of  (\ref{hickory}) in its totality  in a homogeneous magnetic field is  in this approximation by    (\ref{sangrail}) and also (\ref{svanelil}), (\ref{staatrold}) and (\ref{linguini})  found from:
\begin{eqnarray}&&
-\frac{e^2gM_WeB}{8\pi^2}\varepsilon^\mu(k)\varepsilon^\nu(q)
\exp(-\frac{(\vec{k}_\perp+\vec{q}_\perp)^2}{2eB})
\int _0^\infty \tau d\tau e^{-i\tau (M_W^2-eB)}\int _{\frac 12-\epsilon}^{\frac 12+\epsilon} d\alpha  \int_{1-\alpha-\delta}^{1-\alpha}d\gamma
\nonumber\\&& 
\bigg (e^{\delta_{2, {\rm app}}(k, q)}
(  q-i(0, 0, \frac 1B\vec{B}\times \vec{k})
-(1-e^{-2i(1-\alpha-\gamma)\tau eB})(0, 0, \vec{q}_\perp)
-ie^{-2i(1-\alpha-\gamma)\tau eB}(0, 0, \frac 1B\vec{B}\times \vec{q}))_\mu 
\nonumber\\&&
( k+i(0, 0, \frac 1B\vec{B}\times \vec{q})
-(1-e^{-2i(1-\alpha-\gamma)\tau eB})(0, 0, \vec{k}_\perp)
+ie^{-2i(1-\alpha-\gamma)\tau eB}(0, 0, \frac 1B\vec{B}\times \vec{k}))_\nu
\nonumber\\&&
+(k\leftrightarrow q, \mu \leftrightarrow \nu)\bigg)
\label{kalidasa}
\end{eqnarray}
which produces the following  singular terms in addition to those already contained in (\ref{zinfandel}):
\begin{eqnarray}&&
 \frac{e^2gM_W}{4\pi}\varepsilon^\mu(k)\varepsilon^\nu(q)
\exp(-\frac{(\vec{k}_\perp+\vec{q}_\perp)^2}{2eB})
\frac{eB}{\sqrt{(M_H^2+(\vec{k}_\perp+\vec{q}_\perp)^2)(M_W^2-eB-\frac 14 (M_H^2+(\vec{k}_{\perp}+\vec{q}_{\perp})^2))}}
\nonumber\\&&
\bigg(F(0, \theta)
(-q_\mu ((0, 0, \vec{k}_\perp)-i(0, 0, \frac 1B\vec{B}\times \vec{q}))_\nu-((0, 0, \vec{q}_\perp)+i(0, 0, \frac 1B\vec{B}\times \vec{k}))_\mu k_\nu
\nonumber\\&&
+((0, 0, \vec{q}_\perp)+i(0, 0, \frac 1B\vec{B}\times \vec{k}))_\mu ((0, 0, \vec{k}_\perp)-i(0, 0, \frac 1B\vec{B}\times \vec{q}))_\nu)
\nonumber\\&&
+F(1, \theta)
(  q-(0, 0, \vec{q}_\perp)-i(0, 0, \frac 1B\vec{B}\times \vec{k}))_\mu
( (0, 0, \vec{k}_\perp)
+i(0, 0, \frac 1B\vec{B}\times \vec{k}))_\nu
\nonumber\\&&
+(  (0, 0, \vec{q}_\perp)
-i(0, 0, \frac 1B\vec{B}\times \vec{q}))_\mu 
( k-(0, 0, \vec{k}_\perp)+i(0, 0, \frac 1B\vec{B}\times \vec{q}))_\nu)
\nonumber\\&&
+F(2, \theta)
(  (0, 0, \vec{q}_\perp)
-i(0, 0, \frac 1B\vec{B}\times \vec{q}))_\mu ( (0, 0, \vec{k}_\perp)
+i(0, 0, \frac 1B\vec{B}\times \vec{k}))_\nu 
\nonumber\\&&
+(k\leftrightarrow q, \mu \leftrightarrow \nu)\bigg).
\label{sakuntala}
\end{eqnarray}
Also  (\ref{niagara}) is in the same approximation by means of (\ref{sangrail}) combined with (\ref{linguini}), (\ref{bimelech}) and  (\ref{danehill}):
\begin{eqnarray}&&
 \frac{  e^2gM_W}{2\pi}\varepsilon^\mu(k)\varepsilon^\nu(q)\exp(-\frac{(\vec{k}_\perp+\vec{q}_\perp)^2}{2eB})
\frac{e^2B^2}{\sqrt{(M_H^2+(\vec{k}_\perp+\vec{q}_\perp)^2)(M_W^2-eB-\frac 14 (M_H^2+(\vec{k}_{\perp}+\vec{q}_{\perp})^2))}}
\nonumber\\&&
(F(1, \theta) \left( \begin{array}{cc}
 {\bf 0}        &  {\bf 0}\\
{\bf 0} & {\bf 1}-{\bf \sigma}_2
\end{array}\right)_{\mu \nu}
+(\mu \leftrightarrow \nu, k \leftrightarrow q)).
   \label{germanium}
\end{eqnarray}
Finally the singular terms of (\ref{donaulloyd}) that are not included in  (\ref{zinfandel})  are found  by (\ref{hornhyl}) and (\ref{sangrail}) combined with (\ref{linguini}) and (\ref{bimelech}):
\begin{eqnarray}&&
  -\frac{e^2gM_W}{4 \pi}\varepsilon^\mu(k)\varepsilon_\mu(q)
\exp(-\frac{(\vec{k}_\perp+\vec{q}_\perp)^2}{2eB})
\frac{eB}{\sqrt{(M_H^2+(\vec{k}_\perp+\vec{q}_\perp)^2)(M_W^2-eB-\frac 14 (M_H^2+(\vec{k}_{\perp}+\vec{q}_{\perp})^2))}}
\nonumber\\&&
(\vec{q}_{\perp}\cdot\vec{k}_{\perp}F(0, \theta)-\mid\vec{q}_{\perp}\mid \mid\vec{k}_{\perp} \mid e^{-i\theta}F(1, \theta)
+(k \leftrightarrow q)).
\label{lamprey}
\end{eqnarray}

In summary, we have isolated from  (\ref{hickory}), (\ref{niagara}) and (\ref{donaulloyd}) the  terms (\ref{zinfandel}), (\ref{sakuntala}),  (\ref{germanium}) and (\ref{lamprey})  of the $H\rightarrow \gamma \gamma$ amplitude in a homogeneous background magnetic field with the singular factor $\frac{eB}{\sqrt{M_W^2-eB-\frac 14 (M_H^2+(\vec{k}_{\perp}+\vec{q}_{\perp})^2)}}$ and the damping factor $\exp(-\frac{(\vec{k}_\perp+\vec{q}_\perp)^2}{2eB})$. The sum is invariant under  gauge transformations of the polarization vectors; this is demonstrated explicitly in app. C.

Using the second term of the factor $\sin (\tau eB)$ which occurs in the integrands of (\ref{hickory}), (\ref{niagara}) and (\ref{donaulloyd}) in a homogeneous background magnetic field one obtains amplitude terms with the opposite sign and where the square root factor is $\frac{eB}{\sqrt{M_W^2+eB-\frac 14 (M_H^2+(\vec{k}_{\perp}+\vec{q}_{\perp})^2)}}$, cf. the last term of (\ref{beesiana}). From (\ref{frodobaggins}), (\ref{marlash}) and   (\ref{mehemet}), from the remaining parts of  (\ref{laramie}) and  from  (\ref{bedini})  one obtains   also similar amplitude terms with this square root factor.

Defining:
\begin{equation}
 \eta _{\parallel}=(1, -1, 0, 0)
\label{soldat}
\end{equation}
one finds
(\ref{ledreborg})  in a pure magnetic field, 
approximated in the same way as (\ref{wolfgang})-(\ref{zinfandel}) and  using  (\ref{linguini}) and (\ref{bimelech}):
\begin{eqnarray}&&
 -\frac{e^2gM_W}{4\pi}( k^\sigma \varepsilon^\rho(k)-k^\rho \varepsilon^\sigma(k)) (q^\epsilon\varepsilon^\omega(q)-q^\omega\varepsilon^\epsilon(q))\exp(-\frac{(\vec{k}_\perp+\vec{q}_\perp)^2}{2eB})
\nonumber\\&&
\frac{eB}{\sqrt{(M_H^2+(\vec{k}_\perp+\vec{q}_\perp)^2)(M_W^2-eB-\frac 14 (M_H^2+(\vec{k}_{\perp}+\vec{q}_{\perp})^2))}}
\nonumber\\&&
\bigg (\bigg(F(0, \theta)\left( \begin{array}{cc}
{\bf 0}        &  {\bf 0}\\
{\bf 0} & {\bf 1}+ {\bf \sigma}_2
\end{array}
\right)_{\sigma \omega}
\left( \begin{array}{cc}
{\bf 0}        &  {\bf 0}\\
{\bf 0} & {\bf 1}+ {\bf \sigma}_2
\end{array}
\right)_{\epsilon \rho}
+2F(1, \theta)\eta_{\parallel, \sigma \omega}\left( \begin{array}{cc}
{\bf 0}        &  {\bf 0}\\
{\bf 0} & {\bf 1}+ {\bf \sigma}_2
\end{array}
\right)_{\epsilon \rho}\bigg)
\nonumber\\&&
+( k \leftrightarrow q)\bigg).
\label{pennekamp}
\end{eqnarray}
Also, (\ref{stribenlanz}) is approximately by    (\ref{svanelil}), (\ref{staatrold}),  (\ref{linguini}) and (\ref{bimelech}):
\begin{eqnarray}&&
 \frac{e^2gM_W}{4\pi} \varepsilon^\mu(k)\varepsilon^\nu(q)\exp(-\frac{(\vec{k}_\perp+\vec{q}_\perp)^2}{2eB})
\frac{eB}{\sqrt{(M_H^2+(\vec{k}_\perp+\vec{q}_\perp)^2)(M_W^2-eB-\frac 14 (M_H^2+(\vec{k}_{\perp}+\vec{q}_{\perp})^2))}} 
\nonumber\\&&
\bigg(\left( \begin{array}{cc}
{\bf 0}        &  {\bf 0}\\
{\bf 0} & {\bf \sigma}_2
\end{array}
\right)_{\sigma \rho}(\delta ^{\rho}\hspace{0.1 mm}_\mu k^\sigma - \delta ^{\sigma}\hspace{0.1 mm}_\mu k^\rho)
\bigg(F(0, \theta)( k-(0, 0, \vec{k}_\perp)+i(0, 0, \frac 1B\vec{B}\times \vec{q})
)_\nu 
\nonumber\\&&+F(1,\theta)( (0, 0, \vec{k}_\perp)+i(0, 0, \frac 1B\vec{B}\times \vec{k})
)_\nu\bigg)
\nonumber\\&&
-\left( \begin{array}{cc}
{\bf 0}        &  {\bf 0}\\
{\bf 0} & {\bf \sigma}_2
\end{array}
\right)_{\epsilon \omega}(\delta ^{\omega}\hspace{0.1 mm}_\nu q^\epsilon - \delta ^{\epsilon}\hspace{0.1 mm}_\nu q^\omega)
\bigg(F(0,\theta)( q-(0, 0, \vec{q}_\perp)-i(0, 0, \frac 1B\vec{B}\times \vec{k})
)_\mu
\nonumber\\&&
+F(1, \theta)
( (0, 0, \vec{q}_\perp)-i(0, 0, \frac 1B\vec{B}\times \vec{q})
)_\mu\bigg)
+(\mu \leftrightarrow \nu, k \leftrightarrow q)\bigg).
\label{inzaghi}
\end{eqnarray}

The expressions (\ref{pennekamp}) and (\ref{inzaghi}) again have the same singular factor as (\ref{zinfandel});  (\ref{pennekamp}) is manifestly invariant under gauge trnsformations of the polarization vectors, and in app.C it is shown that (\ref{inzaghi}) shares this property.

\section{Quark  contributions}

Quarks are coupled to the Higgs boson and photon fields  through the interaction Lagrangian:
\begin{eqnarray}&&
-Qe{\cal A}_\mu\bar \psi \gamma^\mu \psi -yH\bar \psi \psi
\end{eqnarray}
with $Q=\frac 23, -\frac 13$,  $y$ the Yukawa coupling constant and $\psi$ the quark field,
leading to the Higgs boson  decay  effective action:
\begin{eqnarray}&&
-yQ^2e^2\int d^4x\int d^4y\int d^4xH(x){\cal A}^\mu(y){\cal A}^\nu(z)
\nonumber\\&&
{\rm tr}(<T\psi(x)\bar \psi(y)>\gamma_\mu <T\psi(y)\bar \psi(z)>\gamma _\nu
 <T\psi(z)\bar \psi(x)>).
\label{lamut}
\end{eqnarray}

In an external field  the quark propagator is:
\begin{eqnarray}&&
<T\psi(x)\psi(x')>=<x\mid \frac{i}{i \gamma \cdot D-yv}\mid x'>
\nonumber\\&&
=<x\mid (-\gamma \cdot \Pi+yv)\int _0^\infty d\tau e^{i\tau(\Pi^2+e{\bf F}\cdot \sigma-y^2v^2)}\mid x'>
\label{pantheon}
\end{eqnarray}
with $\gamma_\mu $ the Dirac matrices and:
\begin{equation}
(\gamma \cdot D)^2=D^2-eF^{\mu \nu}\sigma _{\mu \nu}=D^2-e{\bf F}\cdot \sigma; \ \sigma _{\mu \nu}=\frac 14 i[\gamma_\mu, \gamma _\nu] .
\label{pollyanna}
\end{equation}

 (\ref{lamut}) is in the presence of an external field conveniently reformulated by means of the identity:
\begin{eqnarray}&&
{\rm tr}<x\mid H \frac{i}{-\gamma \cdot \Pi -yv+i\epsilon}\gamma \cdot {\cal A} \frac{i}{-\gamma \cdot \Pi -yv+i\epsilon}\gamma \cdot {\cal A}\frac{i}{-\gamma \cdot \Pi -yv+i\epsilon}\mid x>
 \nonumber\\&&
 =-iyv \hspace{1mm}{\rm tr}<x\mid H\frac{i}{(\gamma \cdot \Pi )^2-y^2v^2+i\epsilon}{\cal A}^2\frac{i}{(\gamma \cdot \Pi )^2-y^2v^2+i\epsilon}\mid x>
 \nonumber\\&&
 +yv\hspace{1mm}{\rm tr}<x\mid H\frac{i}{(\gamma \cdot \Pi )^2-y^2v^2+i\delta}\{\gamma \cdot \Pi, \gamma \cdot {\cal A}\}\frac{i}{(\gamma \cdot \Pi)^2 -y^2v^2+i\epsilon} \{\gamma \cdot \Pi, \gamma \cdot {\cal A}\}
 \nonumber\\&&
 \frac{i}{(\gamma \cdot \Pi )^2-y^2v^2+i\epsilon}\mid x>
 \label{daimon}
\end{eqnarray}
where:
\begin{equation}
\{\gamma \cdot \Pi, \gamma \cdot {\cal A}\}= 2{\cal A}^\mu \Pi_\mu- {\cal F}^{\mu \nu} \sigma_{\mu \nu}.
\label{salamander}
\end{equation}
 (\ref{lamut}) is  in this symbolic notation (including a color factor 3):
\begin{equation}
-3yQ^2e^2 
 {\rm tr}<x\mid H\frac{i}{-\gamma \cdot \Pi -yv+i\delta }\gamma \cdot {\cal A} \frac{i}{-\gamma \cdot \Pi -yv+i\delta}\gamma \cdot {\cal A}\frac{i}{-\gamma \cdot \Pi -yv+i\delta}\mid x>
\label{itelmen}
\end{equation}  
and after use of (\ref{daimon}) one gets the quark contribution to the amplitude as the sum of four terms, two of which are:
\begin{eqnarray}&&
6iy^2Q^2e^2v \varepsilon^\mu(k)\varepsilon_\mu(q)
\int d^4xe^{ipx}\int _0^\infty \tau d\tau e^{-i\tau y^2v^2}{\rm tr}(e^{i \tau e{\bf F}\cdot \sigma })
\nonumber\\&&
\int _0^1d\alpha 
<x\mid  e^{i(1-\alpha) \tau\Pi^2}e^{-i(k+q)X}e^{i\alpha \tau\Pi^2}\mid x>
\nonumber\\&&
-
12 y^2Q^2e^2v \varepsilon^\mu(k)\varepsilon^\nu(q)
\int d^4xe^{ipx}\int _0^\infty \tau^2d\tau e^{-i\tau y^2v^2}{\rm tr}(e^{i \tau e{\bf F}\cdot \sigma })\int _0^1d\alpha d\beta d\gamma
 \delta (1-\alpha -\beta -\gamma)
 \nonumber\\&&
(<x\mid e^{i\alpha \tau\Pi^2}e^{-ik\cdot X} \Pi_\mu e^{i\beta \tau\Pi^2} e^{-iq\cdot X}\Pi_\nu e^{i\gamma \tau\Pi^2}\mid x>+(\mu \leftrightarrow \nu, k\leftrightarrow q))
\label{nanai}
\end{eqnarray}
which are  found from (\ref{plausibel}) and (\ref{thorbellinge})  by the replacements $2\lambda e^2v\rightarrow -3y^2Q^2e^2v$ and $e^{-i\tau M_W^2}\rightarrow e^{-i\tau y^2v^2}$  and by insertion of a factor ${\rm tr}(e^{i \tau e{\bf F}\cdot \sigma })$ in the $\tau$-integral. 
The final two terms of the quark contribution to the amplitude  are:
\begin{eqnarray}&&
12y^2Q^2e^2v \varepsilon^\mu(k)\varepsilon^\nu(q)
\int d^4xe^{ipx}\int _0^\infty \tau^2d\tau e^{-i\tau y^2v^2}\int _0^1d\alpha d\beta d\gamma
 \delta (1-\alpha -\beta -\gamma)
 \nonumber\\&&
\bigg( {\rm tr}(e^{i (\alpha +\gamma )\tau e{\bf F}\cdot \sigma }\sigma _{\mu \rho}k^\rho e^{i \beta \tau e{\bf F}\cdot \sigma }\sigma_{\nu \sigma}q^\sigma)<x\mid e^{i\alpha \tau\Pi^2}e^{-ik\cdot X}  e^{i\beta \tau\Pi^2} e^{-iq\cdot X} e^{i\gamma \tau\Pi^2}\mid x>
\nonumber\\&&
+(\mu \leftrightarrow \nu, k \leftrightarrow q)\bigg)
\label{temujin}
\end{eqnarray}
and: 
\begin{eqnarray}&&
12i y^2Q^2e^2v\varepsilon^\mu(k)\varepsilon^\nu(q)
\int d^4xe^{ipx}\int _0^\infty \tau^2d\tau e^{-i\tau y^2v^2}\int _0^1d\alpha d\beta d\gamma
 \delta (1-\alpha -\beta -\gamma)
 \nonumber\\&&
\bigg( {\rm tr}(e^{i \tau e{\bf F}\cdot \sigma }\sigma _{\mu \lambda}k^\lambda )<x\mid e^{i\alpha \tau\Pi^2}e^{-ik\cdot X}  e^{i\beta \tau\Pi^2} \Pi_\nu e^{-iq\cdot X} e^{i\gamma \tau\Pi^2}\mid x>
 \nonumber\\&&
 +{\rm tr}(e^{i \tau e{\bf F}\cdot \sigma }\sigma _{\nu \rho}q^\rho ) <x\mid e^{i\alpha \tau\Pi^2}e^{-ik\cdot X} \Pi_\mu e^{i\beta \tau\Pi^2}  e^{-iq\cdot X} e^{i\gamma \tau\Pi^2}\mid x>)
+(\mu \leftrightarrow \nu, k \leftrightarrow q)\bigg)
\nonumber\\&&
\label{kubilai}
\end{eqnarray}
which are similar to (\ref{ordovician}) and (\ref{wullenwever}) and can be evaluated in the same way.

If the background field is a magnetic field $B$ in the positive 1-direction one estimates the singular behaviour of (\ref{nanai}), (\ref{temujin}) and (\ref{kubilai}) in the same way as  for (\ref{hickory}), (\ref{niagara}), (\ref{donaulloyd}),  (\ref{ledreborg}) and (\ref{stribenlanz}).  In this case one   finds:
\begin{equation}
e^{i\tau e {\bf F}\cdot \sigma}
=\cos (\tau eB)1-\sin (\tau eB)\gamma_2\gamma_3
\label{tipperarytim}
\end{equation}
that should be compared with (\ref{benazir}). Having in (\ref{tipperarytim}) only $\cos (\tau eB)$ and $\sin (\tau eB)$ compared to  $\cos (2\tau eB)$ and $\sin (2\tau eB)$  in (\ref{benazir})
means that taking over the estimates (\ref{zinfandel}), (\ref{sakuntala}), (\ref{germanium}), (\ref{lamprey}), (\ref{pennekamp}) and (\ref{inzaghi})   one finds  no   singularity   of the type found in sec. 4, the square root factor being in this case $\frac{eB}{\sqrt{y^2v^2-\frac 14 (M_H^2+(\vec{k}_{\perp}+\vec{q}_{\perp})^2)}}$.

\section{Higgs boson self energy}

The Higgs boson self energy is given by the effective action:
\begin{equation}
-\frac 12\int d^4x \int d^4yH(x)\Sigma (x- y)H(y).
\label{sciurus}
\end{equation}
The  function $\Sigma (x-y)$  has by   (\ref{kagemucha}) and (\ref{ghostafson})  several terms; we concentrate on:
\begin{equation}
\Sigma (x- y)\simeq -ig^2M_W^2G_{{\rm vec}}\hspace{0.1 mm}^{\mu \nu}(x, y)
G_{{\rm vec},  \nu \mu}(y, x)
\label{koriander}
\end{equation}
where the Feynman gauge  is used. It turns out that (\ref{koriander}) has a similar singularity as the  $H\rightarrow \gamma \gamma$ amplitude, where the singular term is gauge parameter independent.

From (\ref{koriander}) one gets by Fourier transformation and use of  (\ref{dynamide}) and (\ref{boong}):
\begin{eqnarray}&&
\Sigma ( p)=- ig^2M_W^2
\int _0^\infty \tau d\tau e^{-i\tau M_W^2}{\rm tr}(e^{-2\tau e{\bf F}})
\int _0^1d\alpha  e^{ipx}<x\mid e^{i(1-\alpha)\tau \Pi^2}e^{-ip\cdot X}e^{i\alpha \tau \Pi^2 }\mid x>
\nonumber\\&&
=- ig^2M_W^2
\int _0^\infty \tau d\tau e^{-i\tau M_W^2}{\rm tr}(e^{-2\tau e{\bf F}})<x, \tau \mid x, 0> 
\int _0^1d\alpha  e^{\delta _1(\alpha,  p)}.
\label{akvarel}
\end{eqnarray}
The Higgs boson should be on-shell, i.e. $p^2=M_H^2$.
The self energy is evaluated in a constant homogeneous magnetic field along  the positive 1-axis and with the Higgs boson having the momentum component  $\vec{p}_{\perp}$ orthogonal to the magnetic field.
In this particular case  (\ref{akvarel}) is by  (\ref{macaroni}) and (\ref{benazir}):
\begin{equation}
\Sigma (p)=-\frac{g^2M_W^2}{4\pi^2}
\int _0^\infty d\tau \frac{ eB}{\sin(\tau eB)}(1-\sin^2(\tau eB))e^{-i\tau M_W^2}
\int _0^1d\alpha  e^{\delta _1(\alpha,  p)}.
\label{stanovoj}
\end{equation}
With  the Higgs boson momentum  parallel to the magnetic field one isolates in (\ref{stanovoj}):
\begin{eqnarray}&&
-\frac{i}{8\pi^2}g^2M_W^2eB
\int_0^\infty d\tau e^{-i\tau (M_W^2-eB)}\int_0^1d\alpha e^{i\alpha(1-\alpha)\tau M_H^2}
\nonumber\\&&
=
-\frac{1}{4\pi^2}\frac{g^2M_W^2eB}{M_H}
\frac{1}{\sqrt{M_W^2-eB-\frac{1}{4}M_H^2}}\arcsin \frac{M_H}{2\sqrt{M_W^2-eB}}
\label{hagebutter}
\end{eqnarray}
which is  singular  at $eB=M_W^2-\frac{1}{4}M_H^2$.

One can obtain the singularity of (\ref{hagebutter}) also at nonvanishing $\vec{p}_{\perp}$  by means of (\ref{monrepos}), proceeding as in (\ref{wolfgang}) and (\ref{zinfandel}), with $\frac 12-\epsilon<\alpha<\frac 12+\epsilon, 0<\epsilon<<1$:
\begin{eqnarray}&&
\Sigma (p)\simeq -\frac{i}{8\pi^2}g^2M_W^2 eB e^{-\frac{\vec{p}_{\perp}^2}{2eB}}
\int _0^\infty d\tau e^{-i\tau(M_W^2-eB)}\int _{\frac 12-\epsilon}^{\frac 12+\epsilon} d\alpha e^{i\alpha(1-\alpha)\tau (M_H^2+\vec{p}_{\perp}^2)}
\nonumber\\&&
\simeq -\frac{1}{8\pi}g^2M_W^2 eB e^{-\frac{\vec{p}_{\perp}^2}{2eB}}
\frac{1}{\sqrt{(M_H^2+\vec{p}_{\perp}^2)(M_W^2-eB-\frac 14 (M_H^2+\vec{p}_{\perp}^2))}}
\label{viipuri}
\end{eqnarray}
where in the last step  the limiting case  $M_W^2-eB-\frac 14 (M_H^2+\vec{p}_{\perp}^2)\simeq 0$ with $\epsilon$ kept fixed has been considered. (\ref{viipuri}) reduces to (\ref{hagebutter}) in this limit for $\vec{p}_{\perp}$ vanishing, and it has thus been established that  the Higgs boson self energy is singular here. No other contributions to the one-loop Higgs self energy shows this behavior, and neither does the one-loop correction to the Higgs boson field vacuum expectation value. 

\section{Conclusion and comments}

The $H\rightarrow \gamma \gamma$ decay amplitude has been found to have a singularity where  it diverges (see (\ref{zinfandel}), (\ref{sakuntala}),  (\ref{germanium}), (\ref{lamprey}), (\ref{pennekamp}) and (\ref{inzaghi}))  in  a strong stationary and homogeneous magnetic field, and this phenomenon was  shown to be  invariant under gauge transformations of the photon polarization vectors. The singularity was also observed  for the Higgs boson self energy (eq. (\ref{viipuri})), and in both cases it was found to be caused by the unstable mode discussed in \cite{NO}, \cite{JO}.

It would clearly be of interest to investigate whether this behavior of the amplitude also holds in a more realistic situation, where the magnetic field is time-dependent and inhomogeneous with cylindrical symmetry. For such an investigation a gauge-independent regularization method should be formulated, possible by the tools developed in  the present  paper.

{\bf Acknowledgement:} I am grateful to Professor Poul Olesen for giving a very informative seminar on his recent work,  to Professor Per Osland for a helpful conversation, and to  Dr. D. B. Becciolini for preparing the Feynman diagrams using JaxoDraw \cite{Binosi}. Finally I wish to thank an anonymous referee for his constructive criticism and especially for pointing out the relevance of  the paper by Vanyashin and Terentev \cite{Vanyashin}.

\appendix

\section{Reduction of the $H\rightarrow \gamma \gamma$ decay effective action}

\begin{figure}[ht]
\centering
\includegraphics[width=0.75\textwidth]{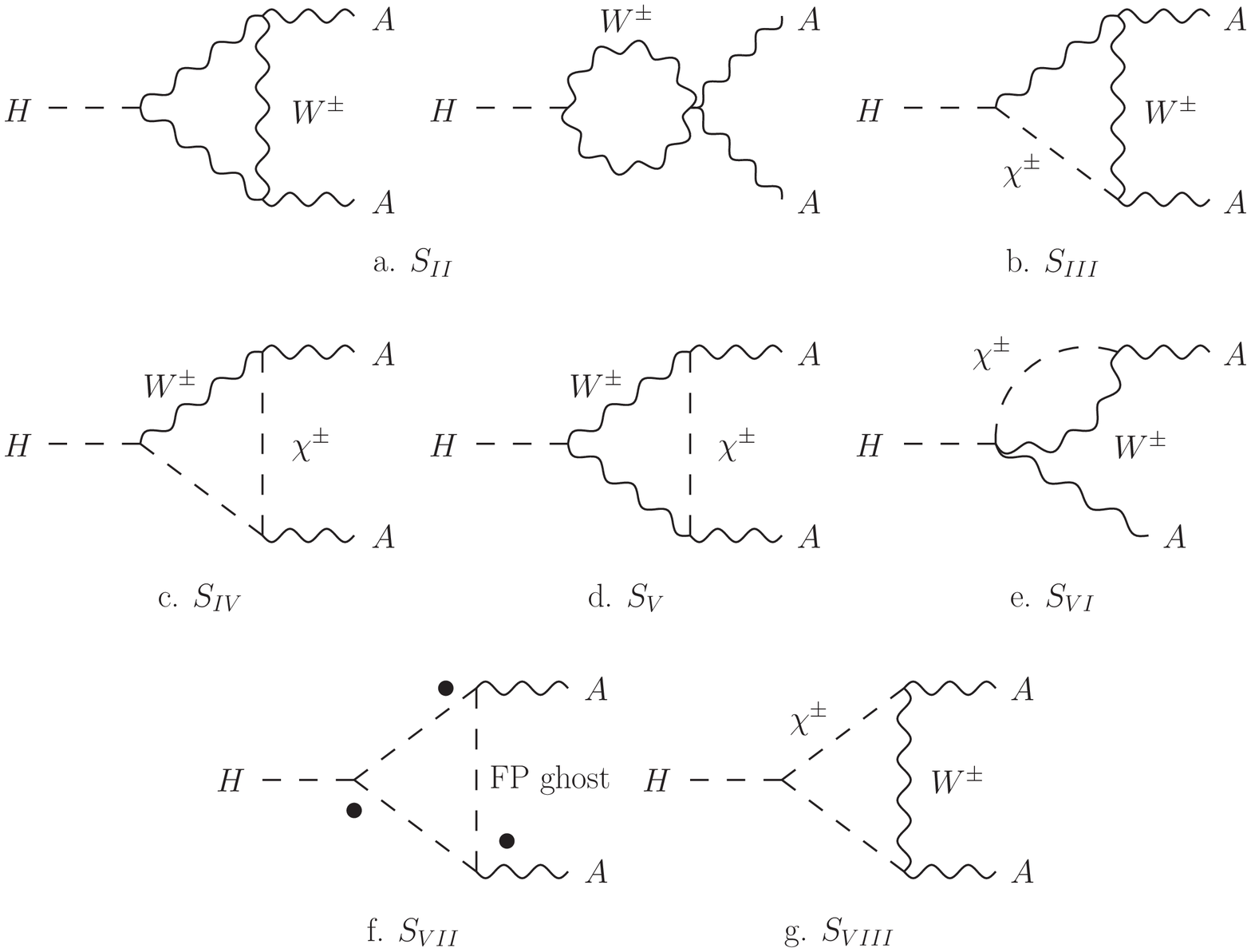}
\caption{Feynman diagram representation of the effective action obtained from (\ref{kagemucha})}
\label{fig:fig2}
\end{figure}

The effective action terms describing Higgs boson decay to two photons are, apart from (\ref{dulwich}):
\begin{eqnarray}&&
S_{II}=
-igM_W\int d^4x\int d^4yH(x)G_{{\rm vec}}\hspace{0.1 mm}^{\mu \nu}(x, y){\cal H}^{(2)}_{\nu \lambda}(y)G_{{\rm vec},}\hspace{0.1 mm}^{\lambda}\hspace{0.1 mm}_\mu (y, x)
\nonumber\\&&
-gM_W\int d^4x\int d^4y\int d^4zH(x)G_{{\rm vec}}\hspace{0.1 mm}^{\mu \nu}(x, y){\cal H}^{(1)}_{\nu \lambda}(y)
G_{{\rm vec}}\hspace{0.1 mm}^{\lambda \rho}(y, z)
{\cal H}^{(1)}_{\rho \sigma}(z)G_{{\rm vec}}\hspace{0.1 mm}^{\sigma}\hspace{0.1 mm}_\mu (z, x),
\nonumber\\&&
\label{arason}
\end{eqnarray}
\begin{eqnarray}&&
S_{III}=-\frac 12iegM_W\int d^4x\int d^4y\int d^4zG_{{\rm vec}}\hspace{0.1 mm}^{\mu \nu}(x, y){\cal H}^{(1)}_{\nu \lambda}(y)
\nonumber\\&&
G_{{\rm vec}}\hspace{0.1 mm}^{\lambda \rho}(y, z){\cal A}_\rho (z) G_{{\rm sc}}(z, x)(\stackrel{\leftarrow}{D}_\mu-\partial _\mu)H(x)
\nonumber\\&&
+\frac 12ie gM_W\int d^4x\int d^4y\int d^4zH(x)(D_\mu-\stackrel{\leftarrow}{\partial}_\mu)G_{{\rm sc}}(x, y)
\nonumber\\&&
{\cal A}_\nu (y) G_{{\rm vec}}\hspace{0.1 mm}^{\nu \lambda }(y, z) {\cal H}^{(1)}_{\lambda \rho}(z)G_{{\rm vec}}\hspace{0.1 mm}^{\rho\mu} (z, x),
\label{ararat}
\end{eqnarray}
\begin{eqnarray}&&
S_{IV}=2e^2 gM_W\int d^4x\int d^4y\int d^4zH(x)(D_\mu-\stackrel{\leftarrow}{\partial}_\mu)G_{{\rm sc}}(x, y)
\nonumber\\&&
{\cal A}^\nu D_\nu G_{{\rm sc}}(y, z) {\cal A}_\lambda (z)
G_{{\rm vec}}\hspace{0.1 mm}^{\lambda \mu}(z, x),
\label{kanchenjunga}
\end{eqnarray}
\begin{eqnarray}&&
S_{V}=-e^2gM_W^3\int d^4x\int d^4y\int d^4zH(x)G_{{\rm vec}}\hspace{0.1 mm}^{\mu \nu}(x, y){\cal A}_{\nu }(y)
G_{{\rm sc}}(y, z)
\nonumber\\&&
{\cal A}^{\lambda }(z)G_{{\rm vec},  \lambda \mu} (z, x),
\label{afzelius}
\end{eqnarray}
\begin{equation}
S_{VI}=ie^2gM_W\int d^4x\int d^4yH(x){\cal A}_\mu(x)G_{{\rm vec}}\hspace{0.1 mm}^{\mu \nu}(x, y){\cal A}_\nu (y) G_{{\rm sc}}( y, x).
\label{agrar}
\end{equation}
There is also  a term of the effective action arising from the Faddev-Popov ghost term (\ref{ghostafson}):
\begin{eqnarray}&&
S_{VII}=e^2 g M_W\int d^4x\int d^4y\int d^4zH(x)G_{{\rm sc}}(x, y){\cal A}^\nu(y)D_\nu G_{{\rm sc}}(y, z)
\nonumber\\&&
{\cal A}^\lambda(y)D_\lambda G_{{\rm sc}}(z, x)
\label{grettir}
\end{eqnarray}
where the ghost propagator was replaced by the Goldstone boson propagator since the masses are equal.
The following  term of the effective action involves the scalar coupling $\lambda$:
\begin{equation}
S_{VIII}=2\lambda  e^2M_W^2v\int d^4x\int d^4y\int d^4z H(x)G_{{\rm sc}}(x, y){\cal A}^\nu(y)G_{{\rm vec},  \nu \lambda}(y, z)
{\cal A}^{ \lambda}(z)G_{\rm {sc}}(z,x),
\label{frederik}
\end{equation}
$S_{II}-S_{VIII}$ have the Feynman diagram representation shown in Figure \ref{fig:fig2}.

The second term of  (\ref{arason}) contains, apart from (\ref{laramie}), (\ref{appalachian}) and (\ref{athapaskian}), two terms that  are reformulated by the Ward identities (\ref{Ward}); they are:
\begin{eqnarray}&&
-ie g M_W \int d^4x\int d^4y\int d^4zH(x) {\cal A}^\mu (y)
\nonumber\\&&
G_{{\rm vec},  \lambda \rho}(x, y)(-\delta ^{\rho}\hspace{0.1 mm}_\mu D^\sigma+\delta ^{\sigma}\hspace{0.1 mm}_\mu \stackrel{\leftarrow}{D}^\rho)G_{{\rm vec},  \sigma \omega }(y, z)
({\cal H}^{(1)})^{\omega \epsilon}(z)G_{{\rm vec}, \epsilon}\hspace{0.1 mm}^\lambda (z, x)
\nonumber\\&&
-ie g M_W \int d^4x\int d^4y\int d^4z H(x){\cal A}^{\nu}(z)
\nonumber\\&&
G_{{\rm vec},  \lambda \rho}(x, y)({\cal H}^{(1)})^{\rho \sigma}(y)G_{{\rm vec},  \sigma \omega }(y, z)(-\delta ^{\omega}\hspace{0.1 mm}_\nu D^\epsilon+\delta ^{\epsilon}\hspace{0.1 mm}_\nu \stackrel{\leftarrow}{D}^\omega)G_{{\rm vec},  \epsilon}\hspace{0.1 mm}^\lambda (z, x)
\label{azkaban}
\end{eqnarray}
and:
\begin{eqnarray}&&
-e^2 g M_W \int d^4x\int d^4y\int d^4zH(x) {\cal A}^\mu (y){\cal A}^{\nu}(z)
\nonumber\\&&
G_{{\rm vec},  \lambda \rho}(x, y)(-\delta ^{\rho}\hspace{0.1 mm}_\mu D^\sigma+\delta ^{\sigma}\hspace{0.1 mm}_\mu \stackrel{\leftarrow}{D}^\rho)G_{{\rm vec},  \sigma \omega }(y, z)
(-\delta ^{\omega}\hspace{0.1 mm}_\nu D^\epsilon+\delta ^{\epsilon}\hspace{0.1 mm}_\nu \stackrel{\leftarrow}{D}^\omega)G_{{\rm vec},  \epsilon}\hspace{0.1 mm}^\lambda (z, x).
\label{navajo}
\end{eqnarray}

(\ref{azkaban}) contains by  (\ref{ehuduanna}):
\begin{eqnarray}&&
-e ^2g M_W \int d^4x\int d^4y\int d^4zH(x) {\cal A}^\mu (y){\cal A}^{\nu}(z)
\nonumber\\&&
(G_{{\rm vec},  \lambda \mu}(x, y) G_{{\rm sc} }(y, z)
(\eta_{\nu \omega}D^2-2ieF_{\nu \omega}(z)-D_{\nu}D_{ \omega})
G_{{\rm vec} }\hspace{0.1 mm}^{\omega \lambda} (z, x)
\nonumber\\&&
+G_{{\rm vec},  \lambda }\hspace{0.1 mm}^{  \rho}(x, y)( \stackrel{\leftarrow}{D}^2\eta_{\rho \mu}-2ie F_{\rho \mu}(y)-\stackrel{\leftarrow}{D}_\rho \stackrel{\leftarrow}{D}_{\ \mu})G_{{\rm sc} }(y,  z)G_{{\rm vec},  \nu}\hspace{0.1 mm}^\lambda (z, x)).
\label{lammefjord}
\end{eqnarray}
The rest of  (\ref{azkaban}) is added to (\ref{ararat}), and the sum is by  (\ref{ehuduanna}) and (\ref{Ward}):
\begin{eqnarray}&&
\frac 12e^2gM_W\int d^4x\int d^4y\int d^4zH(x) G_{{\rm sc}}(x, y){\cal A}^\mu(y)(\eta _{\mu \lambda  }D^2-2ieF_{\mu\lambda}(y)-D _{\mu}D_{ \lambda  })
\nonumber\\&&
G_{{\rm vec}}\hspace{0.1 mm}^{\lambda \nu}(y, z){\cal A}_\nu (z) G_{{\rm sc}}(z, x)
\nonumber\\&&
+\frac 12e ^2gM_W\int d^4x\int d^4y\int d^4zH(x)G_{{\rm sc}}(x, y)
\nonumber\\&&
{\cal A}_\mu (y) G_{{\rm vec}}\hspace{0.1 mm}^{\mu \lambda }(y, z)(\eta _{\lambda \nu }\stackrel{\leftarrow}{D}^2-2ieF_{\lambda\nu}(z)-\stackrel{\leftarrow}{D} _{\lambda }\stackrel{\leftarrow}{D}_{\nu}){\cal A}^\nu (z) G_{{\rm sc}} (z, x).
\label{wimsey}
\end{eqnarray}

(\ref{lammefjord})  is by (\ref{magenta}) and (\ref{Ward}) the sum of:
\begin{equation}
-2ie ^2g M_W \int d^4x\int d^4y H(x) 
{\cal A}^{\mu}(x)G_{{\rm vec},  \mu \nu}(x, y)  {\cal A}^\nu (y)G_{{\rm sc} }(y, x)
\label{millicent}
\end{equation}
and also:
\begin{equation}
2e ^2g M_W^3 \int d^4x\int d^4y\int d^4zH(x)
G_{{\rm vec},  \lambda \mu}(x, y){\cal A}^\mu (y) G_{{\rm sc} }(y, z) {\cal A}^{\nu}(z)
G_{{\rm vec},  \nu }\hspace{0.1 mm}^{ \lambda} (z, x)
\label{lavarello}
\end{equation}
 and:
\begin{eqnarray}&&
e ^2g M_W \int d^4x\int d^4y\int d^4zH(x) {\cal A}^\mu (y){\cal A}^{\nu}(z)
\nonumber\\&&
(G_{{\rm vec},  \lambda \mu}(x, y) G_{{\rm sc} }(y, z)
D_{\nu}
G_{{\rm sc} } (z, x)\stackrel{\leftarrow}{D}^\lambda
\nonumber\\&&
+D_\lambda G_{{\rm sc}}(x, y)  \stackrel{\leftarrow}{D}_{\mu}G_{{\rm sc} }(y,  z)G_{{\rm vec}, \nu}\hspace{0.1 mm}^\lambda (z, x)).
\label{rosenkaal}
\end{eqnarray}
Also (\ref{wimsey}) contains the first term of (\ref{bedini}), as well as:
\begin{equation}
-e ^2gM_W^3\int d^4x\int d^4y\int d^4zH(x)G_{{\rm sc}}(x, y)
{\cal A}_\mu (y) G_{{\rm vec}}\hspace{0.1 mm}^{\mu \nu  }(y, z){\cal A}_\nu (z) G_{{\rm sc}} (z, x)
\label{lacota}
\end{equation}
and also:
\begin{equation}
e^2gM_W\int d^4x\int d^4y\int d^4zH(x) G_{{\rm sc}}(x, y){\cal A}^\mu (y)
D _{\mu}
G_{{\rm sc}}(y, z){\cal A}^\nu (z)D_{ \nu } G_{{\rm sc}}(z, x).
\label{willbury}
\end{equation}

From (\ref{navajo}) one gets, again by  (\ref{Ward}):
\begin{eqnarray}&&
e ^2g M_W \int d^4x\int d^4y\int d^4zH(x) {\cal A}^\mu (y){\cal A}^{\nu}(z)
\nonumber\\&&
(G_{{\rm vec},  \lambda \mu}(x, y)G_{{\rm sc}}(y, x)D_\nu G_{{\rm sc} }( z, x)\stackrel{\leftarrow}{D}^\lambda
\nonumber\\&&
+D_\lambda G_{{\rm sc}}(x, y)\stackrel{\leftarrow}{D}_\mu G_{{\rm sc}}(y, z)G_{{\rm vec}, \nu}\hspace{0.1 mm}^\lambda (z, x))
\label{tampen}
\end{eqnarray}
that  is identical  to (\ref{rosenkaal}), and:
\begin{eqnarray}&&
e ^2g M_W \int d^4x\int d^4y\int d^4zH(x) 
\nonumber\\&&
D_\lambda G_{{\rm sc}}(x, y){\cal A}^\mu (y)G_{{\rm vec},  \mu\nu  }(y, z){\cal A}^{\nu}(z)
G_{{\rm sc}}( z, x)\stackrel{\leftarrow}{D}^\lambda
\label{flores}
\end{eqnarray}
and also:
\begin{eqnarray}&&
e ^2g M_W \int d^4x\int d^4y\int d^4zH(x) 
\nonumber\\&&
G_{{\rm vec},  \lambda \mu}(x, y){\cal A}^\mu (y)D^2G_{{\rm sc}}( y, z){\cal A}^{\nu}(z)G_{{\rm vec},  \nu}\hspace{0.1 mm}^\lambda (z, x).
\label{sumbawa}
\end{eqnarray}

 (\ref{rosenkaal}), (\ref{willbury}) and (\ref{tampen}) are  added to (\ref{kanchenjunga}) and (\ref{grettir}); using again (\ref{Ward})  one obtains the second term of (\ref{bedini}).
Also  (\ref{flores}) is  by the background Higgs boson field on-shell condition and (\ref{carmoisin}) the sum of:
\begin{equation}
-2\lambda e^2M_W^2v \int d^4x\int d^4y\int d^4zH(x)
 G_{{\rm sc}}(x, y) {\cal A}^\mu (y)G_{{\rm vec},  \mu\nu  }(y, z){\cal A}^{\nu}(z)
G_{{\rm sc}}( z, x)
\label{vesteranflod}
\end{equation}
that cancels with (\ref{frederik}),  as well as:
\begin{equation}
 ie ^2g M_W \int d^4x\int d^4yH(x){\cal A}^\mu (x)G_{{\rm vec}, \mu\nu  }(x, y){\cal A}^{\nu}(y)
G_{{\rm sc} }( y, x)
\label{mamrelund}
\end{equation}
and:
\begin{equation}
e ^2g M_W ^3\int d^4x\int d^4y\int d^4zH(x) 
 G_{{\rm sc}}(x, y){\cal A}^\mu (y)G_{{\rm vec},  \mu\nu  }(y, z){\cal A}^{\nu}(z)
G_{{\rm sc} }( z, x)
\label{zuleika}
\end{equation}
that cancels with (\ref{lacota}).
Finally  (\ref{sumbawa}) is by (\ref{carmoisin}) the sum of:
\begin{equation}
-ie ^2g M_W \int d^4x\int d^4yH(x) 
G_{{\rm vec},  \lambda \mu}(x, y){\cal A}^\mu (y){\cal A}^{\nu}(y)G_{{\rm vec},  \nu}\hspace{0.1 mm}^\lambda (y, x)
\label{laurvigen}
\end{equation}
that cancels the remainder of the first term of (\ref{arason}), and:
\begin{equation}
-e ^2g M_W^3 \int d^4x\int d^4y\int d^4zH(x) 
G_{{\rm vec},  \lambda \mu}(x, y){\cal A}^\mu (y)G_{{\rm sc}}( y, z){\cal A}^{\nu}(z)G_{{\rm vec},  \nu}\hspace{0.1 mm}^\lambda (z, x).
\label{wedeljarlsberg}
\end{equation}
 (\ref{millicent}) and (\ref{mamrelund}) cancel with (\ref{agrar}), and (\ref{lavarello})  and (\ref{wedeljarlsberg})  cancel with  (\ref{afzelius}).

In summary  (\ref{arason}), (\ref{ararat}), (\ref{kanchenjunga}), (\ref{afzelius}), (\ref{agrar}), (\ref{frederik}) and (\ref{grettir})  have been reduced to  (\ref{laramie}), (\ref{appalachian}), (\ref{athapaskian}) and (\ref{bedini}) that are invariant under gauge transformations of the radiation field ${\cal A}_\mu$ as shown in app. C.

 Using proper-time regularization one finds  additional terms 
from (\ref{lammefjord}) and   (\ref{sumbawa}) by the methods developed in \cite{NKNny}:
\begin{eqnarray}&&
-ie^2gM_W\int_0^\infty d\tau\frac{\partial}{\partial \tau}\bigg(\tau^2 \int _0^1d\alpha d\beta d\gamma \delta(1-\alpha-\beta-\gamma)
\nonumber\\&&
\int d^4x\int d^4y\int d^4zH(x)
h_{{\rm vec}, \lambda \mu}(x, y;\alpha \tau) {\cal A}^\mu (y)h_{{\rm sc} }(y, z; \beta \tau)
 {\cal A}_{\nu}(z)
h_{{\rm vec},  }\hspace{0.1 mm}^{\nu \lambda} (z, x;\gamma \tau)\bigg)
\nonumber\\&&
\simeq \frac {1}{32\pi^2}e^2gM_W\int d^4x H(x){\cal A}^\mu (x){\cal A}_\mu(x)
\label{labourde}
\end{eqnarray} 
while the  corresponding additional terms from   (\ref{wimsey}) and (\ref{flores})  cancel out. (\ref{labourde}) is not invariant under a gauge transformation of the radiation field ${\cal A}_\mu(x)$ and should be discarded.  It seems to be a general deficiency of the proper-time regularization method that such  expressions occur and should be eliminated either by hand or by use of dimensional regularization \cite{NKNny}.

\section{Propagators and kernels in a  homogeneous background electromagnetic field}

\subsection{The scalar kernel in a homogeneous electromagnetic field}

The starting point for finding propagators in a homogeneous background field is the scalar kernel  determined by Schwinger \cite{Schwinger}:
\begin{equation}
<x, \tau\mid x', 0>=<x\mid e^{-i\tau H}\mid x'>; \  <x, \tau\mid=<x\mid e^{-i\tau H}
\label{kernel}
\end{equation}
with the quasi-Hamiltonian:
\begin{equation}
H=-\Pi^2=-\eta ^{\mu \nu}\Pi_\mu \Pi_\nu
\end{equation}
where
 $\Pi_\mu = -iD_\mu= -i(\partial_\mu-ieA_\mu)$. 
A position operator $X_\mu $ is introduced, with: 
\begin{equation}
X_\mu \mid x>=x_\mu \mid x>
\label{lassalle}
\end{equation}
such that:
\begin{equation}
[\Pi_\mu, X_\nu]=-i\eta_{\mu \nu},
 [X_\mu, X_\nu]=0, \  [\Pi_\mu, \Pi_\nu]=ieF_{\mu \nu}.
\label{pierre}
\end{equation}
The field strength $F_{\mu \nu}$ is assumed homogeneous.

$X_\mu$ and $\Pi_\mu$ can be considered operators in a quasi-Heisenberg picture \cite{Schwinger}. Thus their proper-time development is governed by:
\begin{equation}
\frac{dX_\mu}{d\tau}=-i[X_\mu, H]=-2\Pi_\mu
\end{equation}
and:
\begin{equation}
\frac{d\Pi_\mu}{d\tau}=-i[\Pi_\mu, H]=-2eF_{\mu} \hspace{-0.1 mm} ^{ \nu}\Pi_\nu
\end{equation}
or in a matrix notation:
\begin{equation}
\frac{dX}{d\tau}=-2\Pi, \ \frac{d\Pi}{d\tau}=-2e{\bf F}\Pi
\end{equation}
with solutions:
\begin{equation}
\Pi(\tau)=e^{-2\tau e{\bf F}}\Pi(0), \ X(\tau)=X(0)-{\bf D}(\tau)\Pi(0)
\label{trudeau}
\end{equation}
where:
\begin{equation}
  {\bf D}(\tau)=\frac{{\bf 1}-e^{-2\tau e{\bf F}}}{e{\bf F}}.
\label{doonesbury}
\end{equation}
From  (\ref{pierre}) and (\ref{trudeau}) follows:
\begin{equation}
<x, \tau \mid \Pi_\mu \mid x, 0>=0, \
<x, \tau \mid \Pi_\mu \Pi_\nu\mid x, 0>
=i({\bf D}^{-1}(\tau))_{\mu \nu}<x, \tau \mid  x, 0>.
\label{borkrigel}
\end{equation}

The scalar and vector propagators in the Feynman gauge are, cf. (\ref{ivantaurus}) and (\ref{timian}):
\begin{eqnarray}&&
G_{{\rm sc}}(x, x')=\int _0^\infty d\tau e^{-i\tau M_W^2}<x, \tau\mid x', 0>,
\nonumber\\&&
G_{{\rm vec }, \mu \nu}(x, x')=-\int _0^\infty d\tau e^{-i\tau M_W^2}(\exp(-2\tau  e {\bf  F}))_{\mu \nu}<x, \tau\mid x', 0>
\label{dynamide}
\end{eqnarray}
using a matrix notation for the background field strength. The kernel determined by Schwinger is at coinciding points:
\begin{equation}
<x, \tau\mid x, 0>=-\frac{i}{16 \pi^2\tau^2}\exp(-\frac 12 {\rm tr} \log \frac{\sinh(\tau e{\bf F})}{\tau e{\bf F}}).
\label{meara}
\end{equation}

Also one finds from (\ref{trudeau}):
\begin{equation}
X(\alpha \tau)
=({\bf 1}-{\bf D}(\alpha \tau){\bf D}^{-1}(\tau))X(0)+{\bf D}(\alpha \tau){\bf D}^{-1}(\tau)X(\tau)
\label{uncleduke}
\end{equation}
The  Baker-Campbell-Hausdorff identity:
\begin{equation}
e^{a+b}=e^ae^be^{-\frac 12[a, b]},
\label{CBH}
\end{equation}
which is valid when $[a, b]$ commutes with $a$ and $b$, combined
with  (\ref{uncleduke}), implies \cite{Adler}, \cite{Tsai}:
\begin{eqnarray}&&
\exp(ik\cdot X(\alpha \tau))
=\exp(ik\cdot {\bf D}(\alpha \tau){\bf D}^{-1}(\tau)  X(\tau))
\nonumber\\&&
\exp(ik\cdot (1-{\bf D}(\alpha \tau){\bf D}^{-1}(\tau) ) X(0))
e^{\delta _1(\alpha,  k)}
\label{boong}
\end{eqnarray}
where:
\begin{equation}
\delta _1(\alpha,  k)=\frac 12 ik\cdot {\bf D}(\alpha \tau){\bf D}((1-\alpha) \tau){\bf D}^{-1}(\tau)k.
\label{alicesprings}
\end{equation}

Using again (\ref{CBH}) and (\ref{boong}) one gets:
\begin{eqnarray}&&
\exp(ik\cdot X((1-\alpha)\tau) \exp(iq\cdot X(\gamma\tau ))
\nonumber\\&&
=\exp(iQ\cdot X(\tau))\exp(i(k+q-Q)\cdot X(0))e^{\delta_2(k, q)}
\label{classicgrandcru}
\end{eqnarray}
with:
\begin{eqnarray}&&
\delta _2(k, q)
 =\frac 12 ik\cdot {\bf D}((1-\alpha)\tau){\bf D}(\alpha \tau){\bf D}^{-1}(\tau)k
+\frac 12 iq\cdot {\bf D}((1-\gamma)\tau){\bf D}(\gamma \tau){\bf D}^{-1}(\tau)q
\nonumber\\&&
+i q\cdot  {\bf D}(\alpha\tau){\bf D}(\gamma \tau){\bf D}^{-1}(\tau)k
\label{debeers}
\end{eqnarray}
where:
\begin{eqnarray}&&
\delta _2(k, q)\mid _{\gamma =1-\alpha}=\delta_1(\alpha, k+q).
\label{scovere}
\end{eqnarray}
For vanishing background field one gets:
\begin{equation}
\delta _2(k, q)= i\tau\alpha\gamma M_H^2
\label{octopussy}
\end{equation}
with $k^2=q^2=0, 2kq=M_H^2$.
Also we  have defined:
\begin{equation}
Q=({\bf D}((1-\alpha)\tau){\bf D}^{-1}(\tau))^Tk+({\bf D}(\gamma \tau ){\bf D}^{-1}(\tau))^Tq
\label{mahelia}
\end{equation}
where the superscript $T$ denotes transposed matrix, and with:
\begin{equation}
e^{2\alpha\tau e{\bf F}}Q=k-({\bf D}(\alpha \tau){\bf D}(\tau)^{-1})^Tk+(({\bf D}((\alpha+\gamma)\tau)-{\bf D}(\alpha \tau)){\bf D}(\tau)^{-1})^Tq
\label{llewellyn}
\end{equation}
and:
\begin{equation}
e^{-2\gamma\tau e{\bf F}}(k+q-Q)
=q-{\bf D}(\gamma \tau){\bf D}(\tau)^{-1}q+({\bf D}((\alpha+\gamma)\tau)-{\bf D}(\gamma \tau)){\bf D}(\tau)^{-1}k.
\label{carmarthen}
\end{equation}

\subsection{A pure  magnetic field}

In a pure homogeneous magnetic field $B$, which for simplicity is taken along  the positive $1$-axis, one gets $F_2 \hspace{-0.1 mm}^3=-F_3\hspace{-0.1 mm}^2=-B, \ {\bf F}=-iB{\bf \sigma}_2$, with ${\bf {\sigma}}_2$  the second Pauli matrix, and (\ref{meara})  is here \cite{Schwinger}:
\begin{eqnarray}&&
<x, \tau \mid x,0>=-\frac{i}{16\pi^2\tau^2} \frac{\tau eB}{\sin(\tau eB)}.
\label{macaroni}
\end{eqnarray}
The apparent singularity at $\tau=\frac{n\pi}{eB}, n\epsilon Z$ is spurious since $\tau$ is an integration variable and the integration path can be deformed to run below the real axis or along the negative imaginary axis.

Also one gets here:
\begin{eqnarray}&&
e^{-2\tau e{\bf F}}
=\left( \begin{array}{cc}
{\bf 1}        &  {\bf 0}\\
{\bf 0} & \cos (2\tau eB){\bf 1}+i\sin(2\tau eB) {\bf \sigma}_2
\end{array}
\right)
\label{benazir}
\end{eqnarray}
 From (\ref{benazir}) follows:
\begin{eqnarray}&&
{\bf D}(\tau)= \left( \begin{array}{cc}
2\tau {\bf 1}        &  {\bf 0}\\
{\bf 0} & \frac{1}{eB}(\sin  (2\tau eB){\bf 1}-i(\cos(2\tau eB)-1) {\bf \sigma}_2)
\end{array}
\right).
\label{minimal}
\end{eqnarray} 
One also finds:
\begin{equation}
e^{-2(1-\alpha-\gamma)\tau e{\bf F}}{\bf D}^{-1}(\tau)
=\left( \begin{array}{cc}
\frac{1}{2\tau} {\bf 1}        &  {\bf 0}\\
{\bf 0} & \frac{eB}{2\sin(\tau eB)}(\cos((1-2(\alpha+\gamma)\tau eB){\bf 1}+i\sin ((1-2(\alpha+\gamma))\tau eB)){\bf \sigma}_2)
\end{array}\right)
\label{woodcock}
\end{equation}
and:
\begin{equation}
{\bf D}(\alpha\tau){\bf D}^{-1}(\tau)= \left( \begin{array}{cc}
\alpha {\bf 1}        &  {\bf 0}\\
{\bf 0} &\frac{\sin (\alpha \tau eB)}{\sin(\tau eB)}(\cos((1-\alpha)\tau eB){\bf 1}-i\sin((1-\alpha)\tau eB) {\bf \sigma}_2
\end{array}
\right)
\label{palisander}
\end{equation}
and thus:
\begin{eqnarray}&&
{\bf D}(\gamma \tau){\bf D}(\alpha\tau){\bf D}^{-1}(\tau)
= \left( \begin{array}{cc}
2\alpha \gamma \tau{\bf 1}        &  {\bf 0}\\
{\bf 0} &{\bf X}
\end{array}
\right)
\label{sandalwood}
\end{eqnarray}
where:
\begin{equation}
{\bf X}=\frac{2\sin (\alpha \tau eB)\sin (\gamma \tau eB )}{eB\sin(\tau eB)}(\cos((1-\alpha-\gamma)\tau eB){\bf 1}-i\sin((1-\alpha-\gamma)\tau eB) {\bf \sigma}_2).
\label{sherwood}
\end{equation}

From (\ref{palisander}) follows at $\tau \rightarrow -i\infty, \alpha \simeq \gamma \simeq \frac 12$:
\begin{eqnarray}&&
{\bf D}(\alpha\tau){\bf D}^{-1}(\tau) \simeq \frac 12({\bf 1}-\left( \begin{array}{cc}
 {\bf 0}        &  {\bf 0}\\
{\bf 0} & {\bf \sigma}_2
\end{array}
\right)),
\nonumber\\&&
{\bf D}((\alpha+\gamma)\tau){\bf D}^{-1}(\tau)  \simeq {\bf 1}-\frac 12(1-e^{-2i(1-\alpha-\gamma)\tau eB}) \left( \begin{array}{cc}
 {\bf 0}        &  {\bf 0}\\
{\bf 0} &{\bf 1}+ {\bf \sigma}_2
\end{array}
\right)
\label{mulholland}
\end{eqnarray}
and (\ref{llewellyn}) and (\ref{carmarthen}) are in this limit for a pure magnetic field:
\begin{eqnarray}&&
e^{2\alpha\tau e{\bf F}}Q\simeq \frac 12(k+  q-i(0, 0, \frac 1B\vec{B}\times \vec{k})
-(1-e^{-2i(1-\alpha-\gamma)\tau eB})(0, 0, \vec{q}_\perp)
\nonumber\\&&
-ie^{-2i(1-\alpha-\gamma)\tau eB}(0, 0, \frac 1B\vec{B}\times \vec{q}))
\label{svanelil}
\end{eqnarray}
and:
\begin{eqnarray}&&
e^{-2\gamma\tau e{\bf F}}(k+q-Q)\simeq  \frac 12(k+q+i(0, 0, \frac 1B\vec{B}\times \vec{q})
-(1-e^{-2i(1-\alpha-\gamma)\tau eB})(0, 0, \vec{k}_\perp)
\nonumber\\&&
+ie^{-2i(1-\alpha-\gamma)\tau eB}(0, 0, \frac 1B\vec{B}\times \vec{k})).
\label{staatrold}
\end{eqnarray}
Here the exponentials are kept in their present form; they vanish at $\alpha+\gamma \neq 1, \tau \rightarrow -i\infty$, but are equal to 1 at $\alpha+\gamma=1$.

In the same limit one gets from (\ref{sherwood}):
\begin{eqnarray}&&
{\bf X}\simeq -\frac{i}{2eB}((1+e^{-2i(1-\alpha-\gamma)\tau eB}){\bf 1}+(e^{-2i(1-\alpha-\gamma)\tau eB}-1){\bf \sigma}_2)
\label{scarletknight}
\end{eqnarray}
and thus from (\ref{debeers}):
\begin{eqnarray}&&
\delta _2(k, q)\simeq  i\alpha \gamma \tau(M_H^2+ (\vec{q}_{\perp}+\vec{k}_{\perp})^2)+i(1-\alpha-\gamma)\tau (\alpha \vec{k}_{\perp}^2+\gamma \vec{q}_{\perp}^2)
\nonumber\\&&
-\frac{1}{2 eB }(\vec{q}_{\perp}+\vec{k}_{\perp})^2
-\frac{e^{-i \theta}}{2 eB }(e^{-2i(1-\alpha-\gamma)\tau  eB }-1)\mid  \vec{q}_\perp\mid \mid   \vec{k}_\perp \mid 
\label{rottingham}
\end{eqnarray}
 with $k^2=q^2=0, 2q\cdot k=M_H^2$ and with:
\begin{equation}
\vec{q}_\perp\cdot \vec{k}_\perp=\mid \vec{q}_\perp\mid \mid \vec{k}_\perp\mid\cos \theta, \frac 1B \vec{B}\cdot (\vec{q}\times \vec{k})=\mid \vec{q}_\perp\mid \mid \vec{k}_\perp\mid\sin \theta.
\label{lullubum}
\end{equation}
 where $\vec{k}_\perp$ and $\vec{q}_\perp$ denote the spatial parts of $k$ and $q$ orthogonal to the magnetic field. With the applications in sec. 4 in mind  one can in  (\ref{svanelil}) and (\ref{staatrold})   take $\alpha =\gamma=\frac 12$ in the non-exponential terms in contrast to (\ref{rottingham}).
 Also (\ref{alicesprings}) is in this limit:
\begin{equation}
\delta _1(\alpha, k)\simeq i\alpha (1-\alpha)\tau(k^2+\vec{k}_{\perp}^2)-\frac{\vec{k}_{\perp}^2}{2 eB}.
\label{monrepos}
\end{equation}

At $\tau \rightarrow -i\infty$ one gets from (\ref{macaroni}):
\begin{equation}
<x, \tau \mid x,0>\simeq \frac{1}{8\pi^2 \tau} eB e^{-i\tau  eB}
\label{linguini}
\end{equation}
and (\ref{benazir})  is for $ \tau \rightarrow -i\infty$  approximately.
\begin{equation}
e^{-2\tau e{\bf F}}
\simeq \left( \begin{array}{cc}
{\bf 1}        &  {\bf 0}\\
{\bf 0} & {\bf 0}
\end{array}
\right)+\frac 12 e^{2i\tau eB}\left( \begin{array}{cc}
{\bf 0}        &  {\bf 0}\\
{\bf 0} & {\bf 1}+ {\bf \sigma}_2
\end{array}
\right).
\label{bimelech}
\end{equation}
From (\ref{woodcock}) one finally gets in this approximation:
\begin{eqnarray}&&
e^{-2(1-\alpha-\gamma)\tau e{\bf F}}{\bf D}^{-1}(\tau) \simeq \left( \begin{array}{cc}
\frac{1}{2\tau} {\bf 1}        &  {\bf 0}\\
{\bf 0} & \frac 12i eBe^{-2i(1-\alpha-\gamma)\tau eB}({\bf 1}-{\bf \sigma}_2)
\end{array}\right).
\label{danehill}
\end{eqnarray}

\section{Invariance of the $H\rightarrow \gamma\gamma$   decay amplitude under gauge transformations of the radiation field}

\subsection{A general background field}

After gauge fixing the radiation field ${\cal A}_\mu(x)$ has a residual gauge freedom   under the  gauge transformation ${\cal A}_\mu(x)\rightarrow  {\cal A}_\mu(x)+\partial_\mu \Lambda(x), \ \partial^2 \Lambda(x)=0$. Doing this gauge transformation on  (\ref{dulwich})  one gets at first order in $\Lambda$:
\begin{eqnarray}&&
-4i \lambda e^2 v\int d^4x\int d^4yH(x) G_{{\rm sc}}(x, y)\partial_\nu({\cal A}^\nu(y)\Lambda(y)) G_{\rm {sc}}(y,x)
\nonumber\\&&
-8\lambda e^2v\int d^4x\int d^4y\int d^4zH(x)G_{{\rm sc} }(x, y)(\partial ^\nu\Lambda )(y)
D_\nu G_{{\rm sc}}(y, z)
{\cal A}^\rho(z)
D_\rho G_{{\rm sc}}(z, x)
\nonumber\\&&
-8\lambda e^2v\int d^4x\int d^4y\int d^4zH(x)G_{{\rm sc} }(x, y){\cal A}^\nu(y)
D_\nu G_{{\rm sc} }(y, z)
(\partial ^\rho \Lambda )(z)
D_\rho G_{{\rm sc}}(z, x)
\label{skarphedin}
\end{eqnarray}
that cancel by partial integration and use of (\ref{carmoisin}).
(\ref{laramie}) and (\ref{bedini})    are invariant under  gauge transformations of the radiation field by the same argument. (\ref{appalachian}) is manifestly invariant.  
From (\ref{athapaskian}) one gets by a  gauge transformation:
\begin{eqnarray}&&
2ie^2 g M_W \int d^4x\int d^4y\int d^4z\int d^4pH(p)e^{ipx}\int d^4k {\cal A}^\mu (k)e^{iky}\int d^4q\Lambda (q)e^{iqz}
\nonumber\\&&
G_{{\rm vec}, \lambda \rho}(x, y)(\delta ^{\rho}\hspace{0.1 mm}_\mu k^\sigma - \delta ^{\sigma}\hspace{0.1 mm}_\mu k^\rho)G_{{\rm vec},  \sigma \omega }(y, z)
(D^2-\stackrel{\leftarrow}{D}^2) G_{{\rm vec},  }\hspace{0.1 mm}^{\omega \lambda} (z, x)
\nonumber\\&&
+2ie^2 g M_W \int d^4x\int d^4y\int d^4z\int d^4pH(p)e^{ipx}\int d^4k \Lambda (k)e^{iky}\int d^4q{\cal A}^{\nu}(q)e^{iqz}
\nonumber\\&&
G_{{\rm vec},  \lambda \rho}(x, y(D^2 -\stackrel{\leftarrow}{D}^2)G_{{\rm vec}, }\hspace{0.1 mm}^ {\rho}\hspace{0.1 mm}_{ \omega }(y, z)(\delta ^{\omega}\hspace{0.1 mm}_\nu q^\epsilon - \delta ^{\epsilon}\hspace{0.1 mm}_\nu q^\omega) G_{{\rm vec},  \epsilon}\hspace{0.1 mm}^\lambda (z, x))
\nonumber\\&&
=0
\label{woods}
\end{eqnarray}
by (\ref{magenta}). Using a proper-time representation in
the two last terms of (\ref{skarphedin})   by (\ref{ivantaurus}) one finds that the additional term corresponding to (\ref{labourde}) vanishes in this case.  The additional term from
 (\ref{woods}) also vanishes.

\subsection{Singular terms in a homogeneous magnetic field}

It is not obvious that the sum of the singular terms of the amplitude (\ref{sakuntala}),  (\ref{germanium}) and (\ref{lamprey}) and also the singular term (\ref{inzaghi}) are invariant under  gauge transformations of the photon polarization vectors, and the approximation procedure used to obtain these expressions means that the result of the preceeding subsection does not apply automatically. It is  verified below that  the approximation procedure  indeed respects gauge invariance.

From (\ref{sakuntala}) one first gets through  $\varepsilon ^\mu(k)\rightarrow ik^\mu \Lambda (k)$:
\begin{eqnarray}&&
-\frac{ie^2gM_W}{4\pi}\Lambda(k)\varepsilon^\nu(q)
\exp(-\frac{(\vec{k}_\perp+\vec{q}_\perp)^2}{2eB})
\nonumber\\&&
\frac{eB}{\sqrt{(M_H^2+(\vec{k}_\perp+\vec{q}_\perp)^2)(M_W^2-eB-\frac 14 (M_H^2+(\vec{k}_{\perp}+\vec{q}_{\perp})^2))}}
\nonumber\\&&
((0, 0, \vec{k}_\perp)_\nu+e^{-i\theta}\mid \vec{q}_\perp\mid \mid \vec{k}_\perp\mid F(1, \theta)((0, 0, \vec{k}_\perp)-i(0, 0, \frac 1B\vec{B}\times \vec{q}))_\nu
\nonumber\\&&
+e^{i\theta}\mid \vec{q}_\perp\mid \mid \vec{k}_\perp\mid 
F(1, -\theta)((0, 0, \vec{k}_\perp)+i(0, 0, \frac 1B\vec{B}\times \vec{q}))_\nu)
\label{autolycus}
\end{eqnarray}
by the following identity, which is a consequence of the definition (\ref{armiger}):
\begin{equation}
(q\cdot k +\vec{q}_{\perp}\cdot \vec{k}_{\perp})F(j, \theta)=\frac 12+e^{-i\theta}\mid \vec{q}_\perp\mid \mid \vec{k}_\perp\mid F(j+1, \theta)-2jeBF(j, \theta)
\label{narrgnistor}
\end{equation}
and also:
\begin{eqnarray}&&
\frac{ie^2gM_W}{4\pi}\Lambda(k) k_\nu\varepsilon^\nu(q)\vec{q}_{\perp}\cdot \vec{k}_{\perp}
\exp(-\frac{(\vec{k}_\perp+\vec{q}_\perp)^2}{2eB})
\nonumber\\&&
\frac{eB}{\sqrt{(M_H^2+(\vec{k}_\perp+\vec{q}_\perp)^2)(M_W^2-eB-\frac 14 (M_H^2+(\vec{k}_{\perp}+\vec{q}_{\perp})^2))}}
\nonumber\\&&
(F(0, \theta)+F(0,-\theta)).
\label{getinghonung}
\end{eqnarray}
Using again (\ref{armiger}) one also gets from (\ref{sakuntala}):
\begin{eqnarray}&&
\frac{ie^2gM_WeB}{4\pi}\Lambda(k)\varepsilon^\nu(q)
\exp(-\frac{(\vec{k}_\perp+\vec{q}_\perp)^2}{2eB})
\nonumber\\&&
\frac{1}{\sqrt{(M_H^2+(\vec{k}_\perp+\vec{q}_\perp)^2)(M_W^2-eB-\frac 14 (M_H^2+(\vec{k}_{\perp}+\vec{q}_{\perp})^2))}}
\nonumber\\&&
((0, 0, \vec{k}_\perp)_\nu+e^{-i\theta}\mid \vec{q}_\perp\mid \mid \vec{k}_\perp\mid F(2, \theta)( (0, 0, \vec{k}_\perp)
+i(0, 0, \frac 1B\vec{B}\times \vec{k}))_\nu
\nonumber\\&&
+e^{i\theta}\mid \vec{q}_\perp\mid \mid \vec{k}_\perp\mid F(2, -\theta)( (0, 0, \vec{k}_\perp)
-i(0, 0, \frac 1B\vec{B}\times \vec{k}))_\nu
\nonumber\\&&
-2eBF(1, \theta)( (0, 0, \vec{k}_\perp)
+i(0, 0, \frac 1B\vec{B}\times \vec{k}))_\nu
\nonumber\\&&
-2eBF(1, -\theta)( (0, 0, \vec{k}_\perp)
-i(0, 0, \frac 1B\vec{B}\times \vec{k}))_\nu).
\label{hornbeam}
\end{eqnarray}
and the final terms obtained from (\ref{sakuntala}) are:
\begin{eqnarray}&&
-\frac{ie^2gM_WeB}{4\pi}\mid  \vec{q}_\perp\mid \mid \vec{k}_{\perp}\mid\Lambda(k)\varepsilon^\nu(q)
\exp(-\frac{(\vec{k}_\perp+\vec{q}_\perp)^2}{2eB})
\nonumber\\&&
\frac{1}{\sqrt{(M_H^2+(\vec{k}_\perp+\vec{q}_\perp)^2)(M_W^2-eB-\frac 14 (M_H^2+(\vec{k}_{\perp}+\vec{q}_{\perp})^2))}}
\nonumber\\&&
(e^{-i\theta}F(1, \theta)
( k-(0, 0, \vec{k}_\perp)+i(0, 0, \frac 1B\vec{B}\times \vec{q}))_\nu)
\nonumber\\&&
+ e^{i\theta}F(1, -\theta)
( k-(0, 0, \vec{k}_\perp)-i(0, 0, \frac 1B\vec{B}\times \vec{q}))_\nu)\bigg)
\label{coromandel}
\end{eqnarray}
and also:
\begin{eqnarray}&&
-\frac{ie^2gM_WeB}{4\pi}\Lambda(k)\varepsilon^\nu(q)\mid  \vec{q}_\perp\mid\mid \vec{k}_{\perp}\mid 
\exp(-\frac{(\vec{k}_\perp+\vec{q}_\perp)^2}{2eB})
\nonumber\\&&
\frac{1}{\sqrt{(M_H^2+(\vec{k}_\perp+\vec{q}_\perp)^2)(M_W^2-eB-\frac 14 (M_H^2+(\vec{k}_{\perp}+\vec{q}_{\perp})^2))}}
\nonumber\\&&
(
 e^{-i\theta} F(2, \theta)( (0, 0, \vec{k}_\perp)
+i(0, 0, \frac 1B\vec{B}\times \vec{k}))_\nu 
\nonumber\\&&
+
 e^{i\theta} F(2,-\theta)( (0, 0, \vec{k}_\perp)
-i(0, 0, \frac 1B\vec{B}\times \vec{k}))_\nu \bigg).
\label{wallmart}
\end{eqnarray}

Also one gets from (\ref{germanium}):
\begin{eqnarray}&&
\frac{ i e^2gM_We^2B^2}{2\pi}\Lambda(k)\varepsilon^\nu(q)\exp(-\frac{(\vec{k}_\perp+\vec{q}_\perp)^2}{2eB})
\nonumber\\&&
\frac{1}{\sqrt{(M_H^2+(\vec{k}_\perp+\vec{q}_\perp)^2)(M_W^2-eB-\frac 14 (M_H^2+(\vec{k}_{\perp}+\vec{q}_{\perp})^2))}}
\nonumber\\&&
 (F(1, \theta)((0, 0, \vec{k}_{\perp}) + i(0, 0, \frac 1B\vec{B}\times \vec{k}_{\perp}))_\nu 
\nonumber\\&&
+
F(1, -\theta) ((0, 0, \vec{k}_{\perp}) - i(0, 0, \frac 1B\vec{B}\times \vec{k}_{\perp}))_\nu )
\label{vanadin}
\end{eqnarray}
and
from (\ref{lamprey}):
\begin{eqnarray}&&
 -\frac{ie^2gM_WeB}{4 \pi}\Lambda(k)k_\nu\varepsilon^\nu(q)
\exp(-\frac{(\vec{k}_\perp+\vec{q}_\perp)^2}{2eB})
\nonumber\\&&
\frac{1}{\sqrt{(M_H^2+(\vec{k}_\perp+\vec{q}_\perp)^2)(M_W^2-eB-\frac 14 (M_H^2+(\vec{k}_{\perp}+\vec{q}_{\perp})^2))}}
\nonumber\\&&
(\vec{q}_{\perp}\cdot\vec{k}_{\perp}F(0, \theta)-\mid \vec{q}_{\perp}\mid \mid \vec{k}_{\perp}\mid e^{-i\theta}F(1, \theta)+\vec{q}_{\perp}\cdot\vec{k}_{\perp}F(0, -\theta)-\mid \vec{q}_{\perp}\mid \mid \vec{k}_{\perp}\mid e^{i\theta}F(1, -\theta)).
\nonumber\\&&
\label{ironclad}
\end{eqnarray}
The sum of (\ref{autolycus}) and (\ref{getinghonung})-(\ref{ironclad}) vanishes.

From (\ref{inzaghi}) one  gets through  $\varepsilon ^\mu(k)\rightarrow ik^\mu \Lambda (k)$ by (\ref{narrgnistor}):
\begin{eqnarray}&&
- \frac{ie^2gM_W}{4\pi} \Lambda(k)\varepsilon^\nu(q)\left( \begin{array}{cc}
{\bf 0}        &  {\bf 0}\\
{\bf 0} &  {\bf \sigma}_2
\end{array}
\right)_{\epsilon \omega}(\delta ^{\omega}\hspace{0.1 mm}_\nu q^\epsilon - \delta ^{\epsilon}\hspace{0.1 mm}_\nu q^\omega)\exp(-\frac{(\vec{k}_\perp+\vec{q}_\perp)^2}{2eB})
\nonumber\\&&
\frac{eB}{\sqrt{(M_H^2+(\vec{k}_\perp+\vec{q}_\perp)^2)(M_W^2-eB-\frac 14 (M_H^2+(\vec{k}_{\perp}+\vec{q}_{\perp})^2))}} 
\nonumber\\&&
(( q\cdot k+ \vec{q}_\perp\cdot \vec{k}_\perp)F(0, \theta)-e^{-i\theta}\mid \vec{q}_\perp\mid \mid  \vec{k}_\perp\mid F(1, \theta)
\nonumber\\&&
-( q\cdot k+ \vec{q}_\perp\cdot \vec{k}_\perp)F(0, -\theta)+e^{i\theta}\mid \vec{q}_\perp\mid \mid  \vec{k}_\perp\mid F(1, -\theta))
\nonumber\\&&
=0.
\label{sleipner}
\end{eqnarray}

\section{Heisenberg-Euler amplitude}

The  decay amplitude obtained from the Heisenberg-Euler effective action \cite{Vanyashin} and involving a $W^\pm$-loop can also be found from (\ref{kagemucha}), (\ref{dynamide}) and  (\ref{meara}):
\begin{equation}
gM_W<W^{-\mu}(x)W^+_\mu(x)>
=\frac{igM_W}{16\pi^2} \int _0^\infty \frac{d\tau}{\tau^2}e^{-i\tau M_W^2} {\rm tr}(e^{-2e{\bf F}\tau})\exp(-\frac 12 {\rm tr} \log \frac{\sinh(\tau e{\bf F})}{\tau e{\bf F}})
\label{heisenberg}
\end{equation}
where the field strength {\bf F}, which is assumed homogeneous, is split according to (\ref{sossamon}), with the momentum of the radiation field ${\cal A}$ going to zero, and only terms of second order in ${\cal A}$ are kept.
Introducing \cite{Schwinger}:
\begin{equation}
 {\cal F}=\frac 14 F_{\mu \nu}F^{\mu\nu}, {\cal G}=\frac 18 \epsilon_{\mu \nu \lambda \rho}F_{\mu \nu}F^{\lambda \rho}
\label{lipwig}
\end{equation}
with $\epsilon_{\mu \nu \lambda \rho}$ the standard antisymmetric symbol, and the eigenvalues of the matrix ${\bf F}$:
\begin{equation}
(F^{(1)}, F^{(2)})=\frac{i}{\sqrt{2}}(\sqrt{{\cal F}+i{\cal G}}\pm \sqrt{{\cal F}-i{\cal G}})
\label{broholmer}
\end{equation}
one finds:
\begin{equation}
\exp(-\frac 12 {\rm tr} \log \frac{\sinh(\tau e{\bf F})}{\tau e{\bf F}})=\frac{\tau eF^{(1)}}{\sinh(\tau eF^{(1)})}\frac{\tau eF^{(2)}}{\sinh(\tau eF^{(2)})}
\label{shadelandlamott}
\end{equation}
and:
\begin{equation}
{\rm tr}(e^{-2e{\bf F}\tau})=2\cosh(2\tau eF^{(1)})+2\cosh(2\tau eF^{(2)})
\label{beautaw}
\end{equation}

With the background field a homogeneous magnetic field and with the photons emitted along the field lines the quantity ${\cal G}$ vanishes also after the  splitting  (\ref{sossamon}), and ${\cal F}$ will not contain terms where the radiation field multiplies the background field (this will not be the case for general directions of emission). Inserting (\ref{shadelandlamott}) and (\ref{beautaw}) into (\ref{heisenberg}) one gets:
\begin{equation}
 \frac{igM_W}{4\pi^2} \int _0^\infty \frac{d\tau}{\tau^2}e^{-i\tau M_W^2} \frac{\tau e\sqrt{2{\cal F}}}{\sin(\tau e\sqrt{2{\cal F}})}(1-\sin^2(\tau e\sqrt{2{\cal F}})).
\label{gulhyndy}
\end{equation}
This expression gets through the splitting ${\cal F}\rightarrow {\cal F}+\delta {\cal F}$ the additional terms at first order in $\delta {\cal F}$:
\begin{equation}
\frac{igM_W}{8\pi^2} \frac{\delta {\cal F}}{{\cal F}}\int _0^\infty \frac{d\tau}{\tau^2}e^{-i\tau M_W^2} \frac{\tau e\sqrt{2{\cal F}}}{\sin(\tau e\sqrt{2{\cal F}})}(1-\tau e\sqrt{2{\cal F}}\cot (\tau e\sqrt{2{\cal F}})) )
\label{khwarezmia}
\end{equation}
and also:
\begin{equation}
- \frac{igM_W}{8\pi^2} \frac{\delta {\cal F}}{{\cal F}}\int _0^\infty \frac{d\tau}{\tau^2}e^{-i\tau M_W^2} \tau e\sqrt{2{\cal F}}\sin(\tau e\sqrt{2{\cal F}})(1-\tau e\sqrt{2{\cal F}}\cot (\tau e\sqrt{2{\cal F}})) )
\label{shahriar}
\end{equation}
and:
\begin{equation}
- \frac{igM_W}{4\pi^2} \frac{\delta {\cal F}}{{\cal F}}\int _0^\infty \frac{d\tau}{\tau^2}e^{-i\tau M_W^2} (\tau e\sqrt{2{\cal F}})^2\cos(\tau e\sqrt{2{\cal F}}).
\label{sheherezade}
\end{equation}

Only (\ref{shahriar}) and (\ref{sheherezade}) are affected by the quasi-tachyonic field component.
They  are compared with the relevant part of the decay amplitude determined previously in the limit where the photon momenta and thus the Higgs boson mass go to zero with the photons emitted along the field lines. The polarization vectors are orthogonal to the field lines in this case. Then it follows from (\ref{llewellyn}) and (\ref{carmarthen}) combined with (\ref{minimal})  that 
(\ref{hickory}) vanishes, while (\ref{niagara}) is by  (\ref{debeers}) with (\ref{minimal}) as well as (\ref{macaroni}),  (\ref{benazir}) and (\ref{woodcock}):
\begin{eqnarray}&&
-\frac{e^3gM_WB}{\pi^2}\varepsilon^\mu(k)\varepsilon_\mu(q)
\int _0^\infty \tau d\tau e^{-i\tau M_W^2}\sin (\tau eB)\int _0^1d\alpha d\beta d\gamma
 \delta (1-\alpha -\beta -\gamma) e^{i\tau \alpha \gamma M_H^2}
\nonumber\\&&
 (\frac{eB\cos((1-2\beta)\tau eB)}{\sin(\tau eB)}-\frac{1}{\tau})
\label{alphabetsoup}
\end{eqnarray}
that at lowest nontrivial order in $M_H^2$  is:
\begin{eqnarray}&&
-\frac{ie^4gM_WB^2}{\pi^2}M_H^2\varepsilon^\mu(k)\varepsilon_\mu(q)\int _0^\infty \tau^2 d\tau e^{-i\tau M_W^2}
\nonumber\\&&
(\frac{1}{24}\frac{1}{\tau eB}\sin (\tau eB)+\frac 18\frac{1}{(\tau eB)^3}(\tau eB\cos(\tau eB)-\sin(\tau eB)))
\label{roidesanimaux}
\end{eqnarray}
which when added to (\ref{maximilian})  is precisely  (\ref{shahriar}) for this particular case.

From (\ref{ledreborg})  one gets in the same limit by (\ref{macaroni}) and (\ref{benazir}):
\begin{eqnarray}&&
\frac{ie^2gM_W}{4\pi^2}\int _0^\infty d\tau  e^{-i\tau M_W^2}\frac{\tau eB}{\sin(\tau eB)}\int _0^1d \beta (1-\beta)
\nonumber\\&&
\bigg ( \left( \begin{array}{cc}
{\bf 1}        &  {\bf 0}\\
{\bf 0} & \cos (2(1-\beta)\tau eB){\bf 1}+i\sin(2(1-\beta)\tau eB) {\bf \sigma}_2
\end{array}\right )_{\epsilon \rho}{\cal F}^{\rho\sigma}(k)
\nonumber\\&&
\left( \begin{array}{cc}
{\bf 1}        &  {\bf 0}\\
{\bf 0} & \cos (2(\beta\tau eB){\bf 1}+i\sin(2\beta\tau eB) {\bf \sigma}_2
\end{array}\right )_{\sigma \omega}{\cal F}^{\omega\epsilon}(q)
\nonumber\\&&
+\left( \begin{array}{cc}
{\bf 1}        &  {\bf 0}\\
{\bf 0} & \cos (2(1-\beta)\tau eB){\bf 1}+i\sin(2(1-\beta)\tau eB) {\bf \sigma}_2
\end{array}\right )_{\epsilon \rho}{\cal F}^{\rho\sigma}(q)
\nonumber\\&&
\left( \begin{array}{cc}
{\bf 1}        &  {\bf 0}\\
{\bf 0} & \cos (2(\beta\tau eB){\bf 1}+i\sin(2\beta\tau eB) {\bf \sigma}_2
\end{array}\right )_{\sigma \omega}{\cal F}^{\omega\epsilon}(k)\bigg)
\label{screwdriver}
\end{eqnarray}
using the Fourier transform of the radiation field strength (\ref{tajmahal}).
With the photons emitted along the field lines and their polarization vectors thus orthogonal to the field lines (\ref{screwdriver}) reduces to:
\begin{eqnarray}&&
\frac{ie^2gM_W}{4\pi^2}\int _0^\infty d\tau  e^{-i\tau M_W^2}
\cos ( \tau eB)
\nonumber\\&&
\bigg ({\cal F}^{\omega\epsilon}(k) \left( \begin{array}{cc}
{\bf 1}        &  {\bf 0}\\
{\bf 0} & {\bf 0}
\end{array}\right )_{\epsilon \rho}{\cal F}^{\rho\sigma}(q)
\left( \begin{array}{cc}
{\bf 0}        &  {\bf 0}\\
{\bf 0} & {\bf 1}
\end{array}\right )_{\sigma \omega}
+{\cal F}^{\omega\epsilon}(q) \left( \begin{array}{cc}
{\bf 1}        &  {\bf 0}\\
{\bf 0} & {\bf 0}
\end{array}\right )_{\epsilon \rho}{\cal F}^{\rho\sigma}(k)
\left( \begin{array}{cc}
{\bf 0}        &  {\bf 0}\\
{\bf 0} &{\bf 1}
\end{array}\right )_{\sigma \omega}\bigg)
\nonumber\\&&
\label{wallbanger}
\end{eqnarray}
which is a special case of (\ref{sheherezade}).

The square-root singularity of (\ref{manderley}) is not obtained from (\ref{shahriar}) or (\ref{sheherezade}).

\end{document}